\documentclass[conference]{IEEEtran}
\usepackage{cite}
\usepackage{amsmath,amssymb,amsfonts}
\usepackage{algorithmic}
\usepackage{graphicx}
\usepackage{textcomp}
\usepackage{xcolor}
\usepackage{fancyhdr}
\usepackage[hyphens]{url}
\usepackage{diagbox}
\usepackage{hyperref}
\hypersetup{
    colorlinks=true,
    linkcolor=green,
    filecolor=magenta,      
    urlcolor=blue,
}
\usepackage[caption=false]{subfig}
\usepackage{balance}

\def\BibTeX{{\rm B\kern-.05em{\sc i\kern-.025em b}\kern-.08em
    T\kern-.1667em\lower.7ex\hbox{E}\kern-.125emX}}

\fancypagestyle{firstpage}{%
  \fancyhf{}%
  \fancyfoot[C]{\thepage}%
}  
\title{Deep Learning Training in Facebook Data Centers: Design of Scale-up and Scale-out Systems}

\author{Maxim Naumov$^*$, John Kim$^\dagger$, Dheevatsa Mudigere$^\ddagger$, Srinivas Sridharan, Xiaodong Wang, \\ Whitney Zhao, Serhat Yilmaz, Changkyu Kim, Hector Yuen, Mustafa Ozdal, Krishnakumar Nair, \\ Isabel Gao, Bor-Yiing Su, Jiyan Yang and  Mikhail Smelyanskiy \\ Facebook, 1 Hacker Way, Menlo Park, CA}

\begin{document}
\maketitle
\thispagestyle{firstpage}
\pagestyle{plain}

\makeatletter{\renewcommand*{\@makefnmark}{}\footnotetext{\noindent $^*$mnaumov@fb.com $^\ddagger$dheevatsa@fb.com \\ $\phantom{..}$ $^\dagger$jjk12@kaist.edu Currently at Korea Advanced Institute of Science and Technology (KAIST). Work done while at Facebook. } \makeatother}

\begin{abstract}

Large-scale training is important to ensure high performance and accuracy of machine-learning models. At Facebook we use many different models, including computer vision, video and language models. However, in this paper we focus on the deep learning recommendation models (DLRMs), which are responsible for more than 50\% of the training demand in our data centers. Recommendation models present unique challenges in training because they exercise not only compute but also memory capacity as well as memory and network bandwidth. As model size and complexity increase, efficiently scaling training becomes a challenge. To address it  we design Zion – Facebook’s next-generation large-memory training platform that consists of both CPUs and accelerators. Also, we discuss the design requirements of future scale-out training systems.

\end{abstract}

\section{Introduction}

Artificial intelligence (AI) applications are rapidly evolving and increasing  the  demands  on  hardware  and  systems.  Machine learning (ML), deep learning (DL) in particular,  has  been  one  of  the  driving  forces  behind  the remarkable progress in AI and has become one of the most demanding  workloads  in  terms  of  compute  infrastructure in the data centers~\cite{Alibaba-PAI,covington2016deep,johnson2016google,borisyuk2018rosetta}. Moreover, the  continued  growth  of  DL  models  in terms  of  complexity,  coupled  with  significant  slowdown  in transistor scaling, has necessitated going beyond traditional general-purpose processors and developing specialized hardware with holistic  system-level  solutions  to  improve performance, power, and efficiency~\cite{TPUtutorial,HuaweiTraining}.

Within Facebook, DL is used across many social network services, including computer vision, i.e. image classification, object detection, as well as video understanding. In addition, it is used for natural language processing, i.e. translation and content understanding.  However, some of the most important DL models within Facebook are the recommendation models used for ranking and click through rate (CTR) prediction, including News Feed and search services~\cite{hazelwood2018applied}. 

The use of DL models is often split into inference and training work categories~\cite{Jiang2019,PaddlePaddle2019}. Details of inference at Facebook has been discussed earlier~\cite{FBinference,Gupta2019}; in comparison, we address the challenges in training and in particular, the scale-out requirements of the deep learning recommendation models (DLRMs) at Facebook~\cite{DLRM}. 

\begin{figure}[t]
  \center 
  \subfloat[]{\includegraphics[width=0.9\columnwidth]{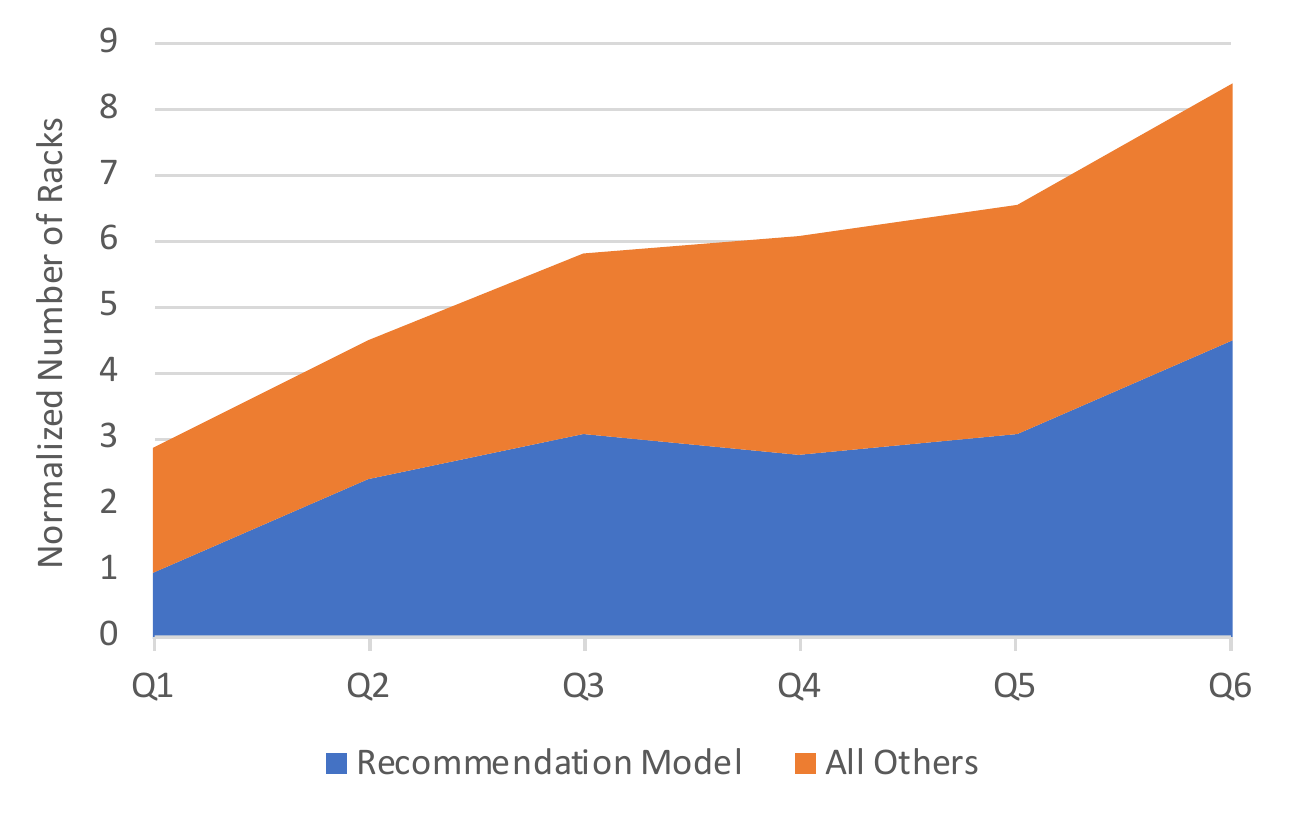}} \hfill \vspace{-12pt}
  \subfloat[]{\includegraphics[width=0.9\columnwidth]{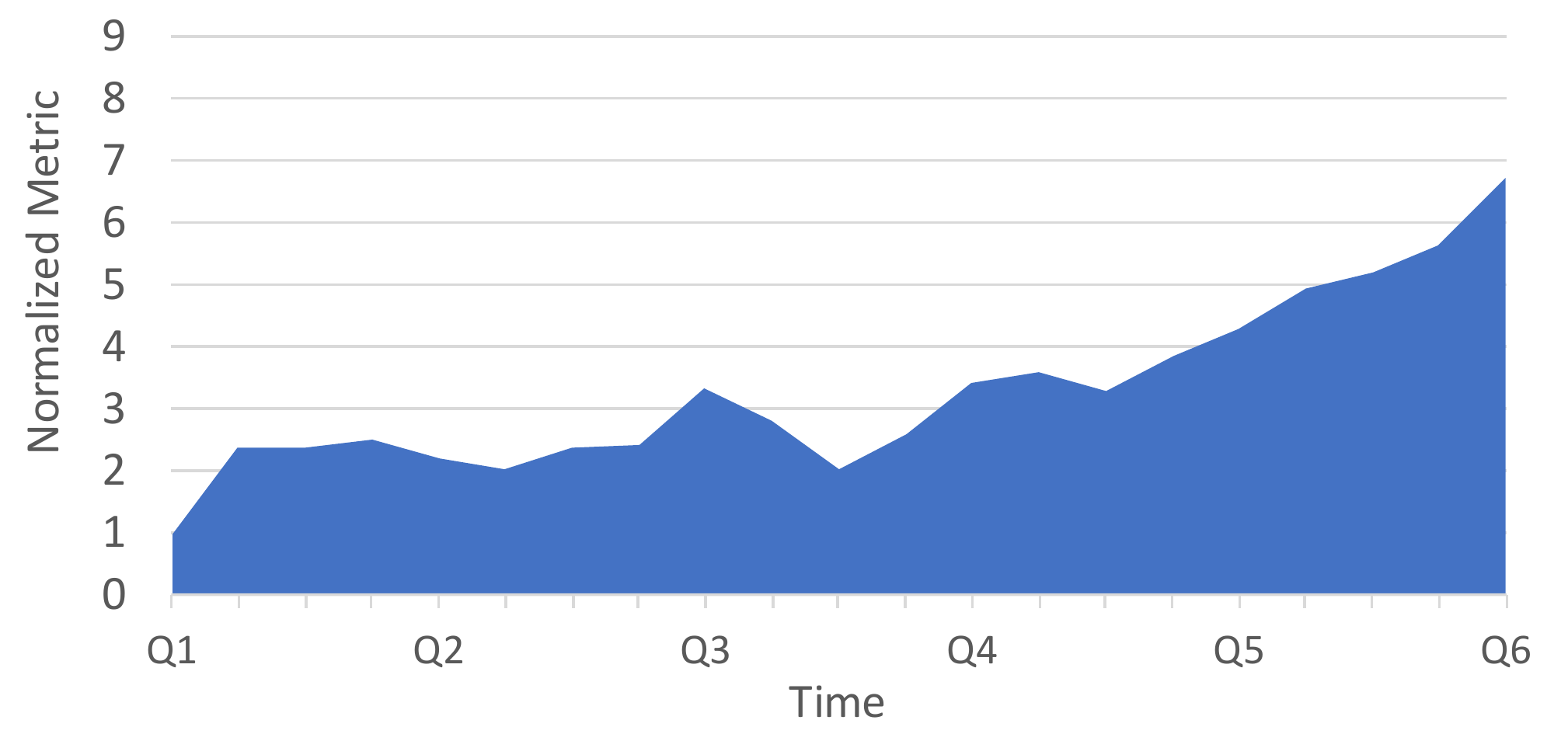}} \hfill \vspace{-5pt}
  \caption{(a) Server compute demand for training and (b) number of distributed training workflows across Facebook data centers.}
  \label{fig:training_trend}
  \vspace{-14pt}
\end{figure}

The increase in compute in Facebook data centers from training is shown in Figure~\ref{fig:training_trend}(a).  Across a period of 18 months, there has been over 4$\times$ increase in the amount of computer resources utilized by training workloads.  In addition, the number of workloads submitted for distributed training, as shown in Figure~\ref{fig:training_trend}(b), has increased at an even higher rate -- resulting in up to 7$\times$ increase in the number of training workflows.  Thus, the demand for training deep learning workloads is continuing to increase while the compute necessary to support it is also increasing proportionally. 

Prior training platform from Facebook, e.g. Big Basin~\cite{BigBasin2019}, consisted of NVidia GPUs. However, it did not leverage other accelerators and only had support for a limited number of CPUs. On the other hand, the Zion next-generation training platform incorporates 8 CPU sockets, with a modular design having separate sub-components for CPUs (Angels Landing 8-socket system) and accelerators (Emeralds Pools 8-accelerator system). This provides for sufficient general purpose compute and more importantly, additional memory capacity. 

Zion also introduced the common form factor OCP Accelerator Module (OAM)~\cite{oam_initiative}\footnote{Proposed and developed as part of the \href{https://www.opencompute.org}{Open Compute Project} (OCP).}, which has been adopted by leading GPU vendors such as NVidia, AMD, Intel, as well as startups, such as Habana which was recently acquired by Intel. This is important for enabling consumers such Facebook to build vendor agnostic accelerator based systems.

In this work, we provide an overview of the DLRM workloads, including the description and analysis of

\begin{itemize}
    \item Training at Facebook
    \item Zion hardware platform
    \item Impact on the accelerator fabric design 
    \item Implications for future scale-out systems
\end{itemize}

\section{Background}

\subsection{Recommendation Model}

\begin{figure}[t]
\center 
 \includegraphics[width=0.95\columnwidth]{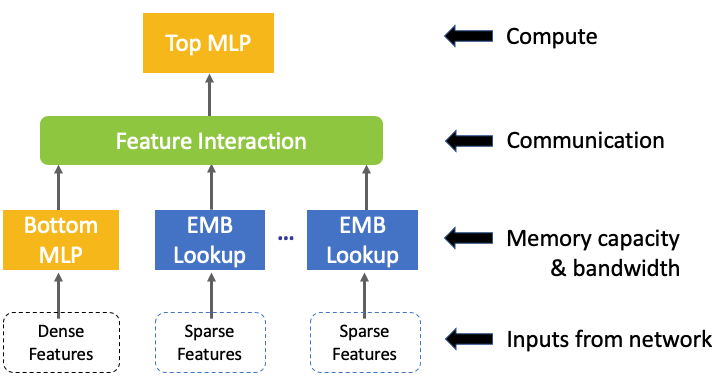}
 \caption{High-level overview of DLRM.}
 \label{fig:dlrm}
\end{figure}

Neural network-based recommendation models, which are used to address personalization for different services, have become an important class of DL algorithms within Facebook. High-level block overview of a typical recommendation model is shown in Figure~\ref{fig:dlrm}, while its implementation in PyTorch framework has been publicly released in DLRM~\cite{DLRM}. 

The inputs to the recommendation model include both dense and sparse features. The dense or the continuous features are processed with a bottom multilayer perceptron (MLP) while the sparse or the categorical features are processed using embeddings. The second-order interactions of different features are computed explicitly. Finally, the results are processed with a top MLP and fed into a sigmoid function in order to provide a probability of a click.

\begin{figure}[t]
  \center 
  \includegraphics[width=0.9\columnwidth]{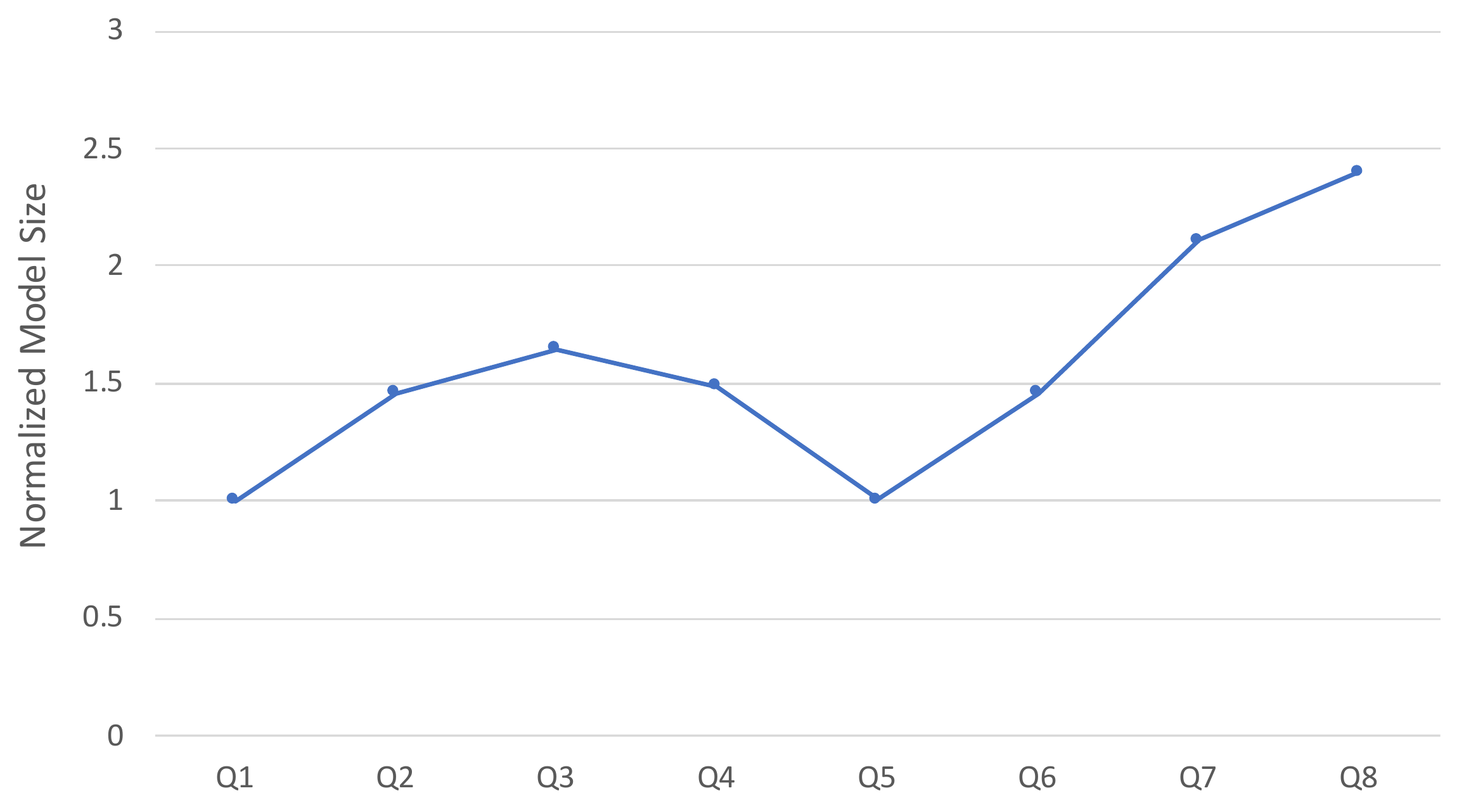}
  \caption{Increase in model complexity over time.}
  \label{fig:model_size}
\end{figure}

\begin{figure}[t]
\center 
\includegraphics[width=0.9\columnwidth]{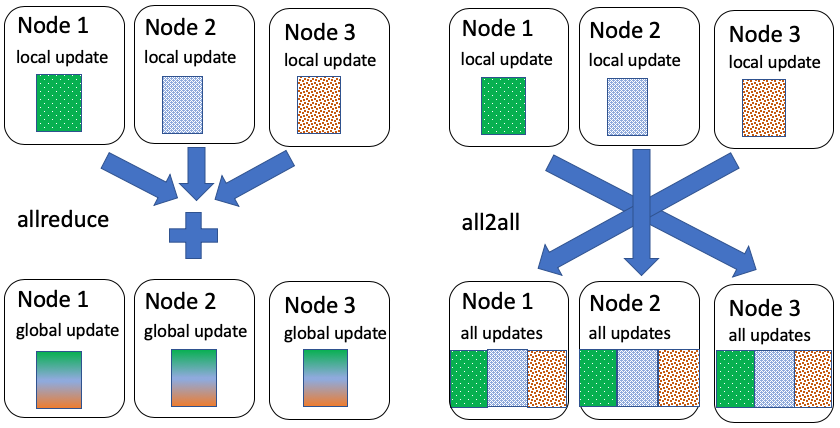}
\centerline{\small{(a)} \hspace{1.5in} \small{(b)}}
\caption{Communication patterns that are common in (a) data- and (b) model-parallelism.  Both communication patterns  need to be supported in DLRM.}
\label{fig:comm_pattern}
\end{figure}

\subsection{Increase in Complexity}

In general the model complexity in terms of number of parameters increases by more than $2\times$ over 2 years, as shown on Figure \ref{fig:model_size}. Notice that the increase trend is not monotonic because in order to alleviate pressure on the training,  inference and serving resources, we are constantly evaluating novel techniques to improve efficiency, such as quantization and compression. However, over time the newly available complexity budget is often reused to improve and augment the more efficient model, therefore driving its size up again. 

\subsection{Distributed Training}
\label{subsec:parallelism}

Distributed training becomes particularly important in the context of increasing model sizes. It may rely on two types of parallelism: data- and model-parallelism. The former enables faster training as each worker trains on different data while the latter enables bigger model to train on the same data. 

\noindent
{\bf In data-parallelism}:  The input data samples are  distributed  across different nodes. Each node processes the input data independently  with  replicas  of  parameters  on  each node,  and  aggregating  the  local  parameter  updates into  a  global  update  on  all  the  nodes.  This  requires  communicating  only  the  updates between the nodes,  but  communication  volume  increases due  to replication as the number of nodes increases. Therefore, scaling out requires large enough  mini-batch  size to provide sufficient parallelism and computation to hide the communication overhead.

\noindent
{\bf In model-parallelism:}  The model weights corresponding to neural network layers are distributed  across  multiple  nodes. Each node processes the entire mini-batch of data and communicates the activations forward or error gradients backwards to other nodes. This introduces additional synchronization across all nodes after each distributed layer in the forward and backward pass. However, it allows us to fit the model into the aggregate memory of all distributed nodes. 

Note that a single embedding table contains tens of millions of vectors, each with hundreds of elements. It requires significant memory capacity, on the order of GBs. Therefore, an embedding table often exists as a single instance and can not be replicated on multiple devices or nodes. In contrast MLPs are relatively small and can be replicated many times. Therefore, we will leverage both types of parallelism while training DLRMs.

\section{Training at Facebook}

\begin{figure}[t]
  \center 
  \includegraphics[width=0.95\columnwidth]{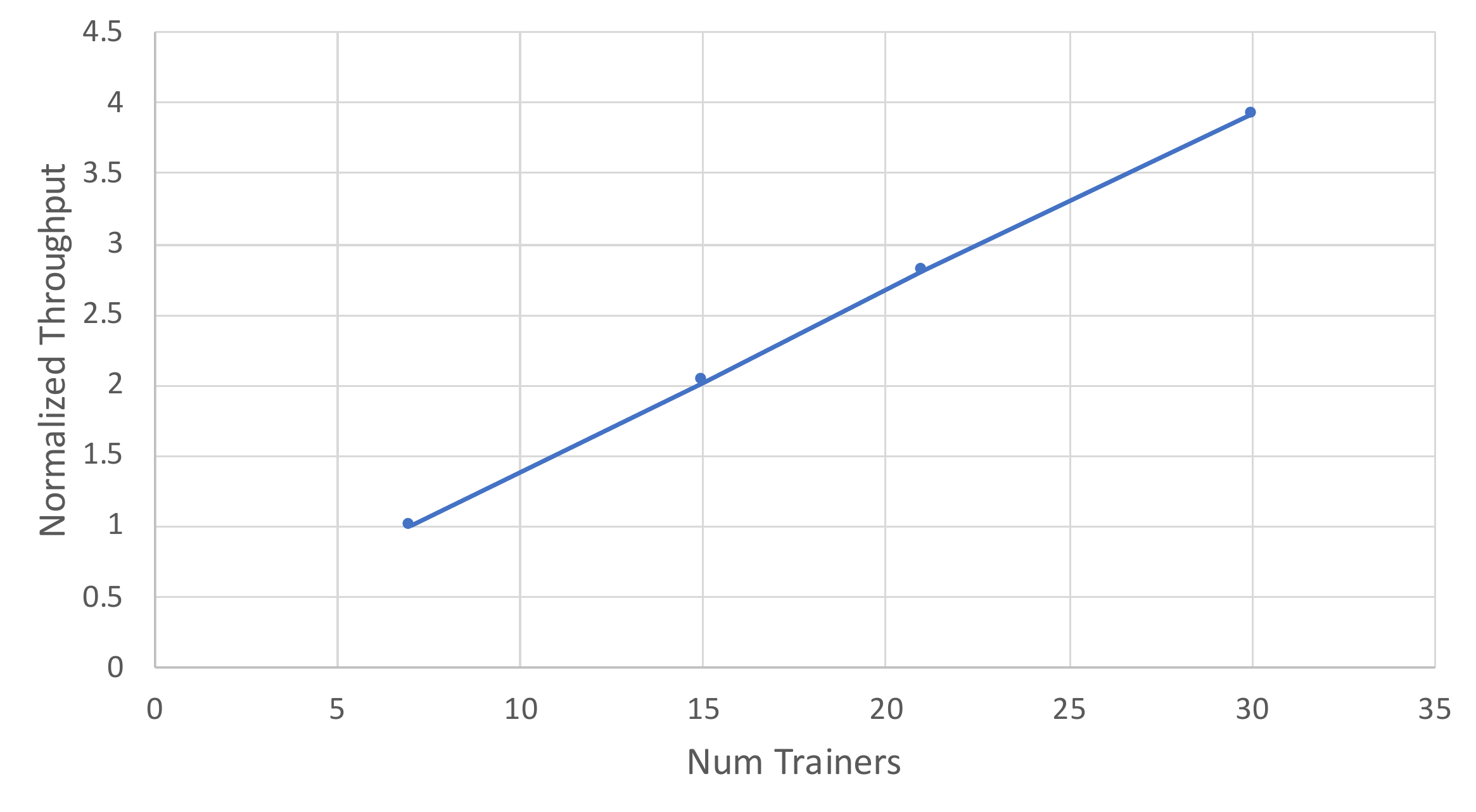}
  \caption{Performance as the number of trainers are increased.}
  \label{fig:trainer}
\end{figure}

\subsection{Overview of Training at Facebook}

The training of recommendation models often requires the distribution of the model across multiple devices within a single node or multiple nodes. Hence requiring both data- and model-parallelism to scale the performance of training~\cite{DLRM}. The distributed training can be performed using a combination of synchronous algorithms that produce results equivalent to a sequential run of the model ~\cite{das2016distributed,googRevisitingSync} or asynchronous algorithms that scale to a larger number of nodes~\cite{Dean2012,Sergeev2018,Zheng2020}. In general, the asynchronous algorithms can perform a single step of forward and backward propagation faster, but may require more steps to achieve convergence or even converge to a sub-optimal minimum~\cite{Zinkevich2010,bottou2016optimization}. 

{\bf Synchronous training} relies on collective communication to achieve the model and data parallelism. 

The compute intensive MLPs are replicated across devices and work on parts of the mini-batch data samples. Notice that training of MLPs requires an \texttt{allreduce} communication primitive to synchronize their weights during backward propagation, as shown in Figure~\ref{fig:comm_pattern}(a). 

Further, a model may contain tens of embedding tables, which can not be replicated due to memory capacity constraints. These tables are often distributed across devices and each of them processes an entire mini-batch of lookups. Then, an embedding lookup produces several vectors corresponding to the elements in the mini-batch. Let us use the same color to denote vectors resulting from a single embedding table. The need to exchange these vectors in the forward pass and their gradients in the backward pass gives rise to the \texttt{alltoall} communication primitive, as shown in Figure~\ref{fig:comm_pattern}(b).

In this setting, embedding lookups take advantage of the aggregate memory bandwidth across devices, while the full model exercises the interconnect, because multiple lookup results need to be communicated through to be exchanged and computed at once. 

\begin{figure}[t]
  \center 
  \includegraphics[width=0.95\columnwidth]{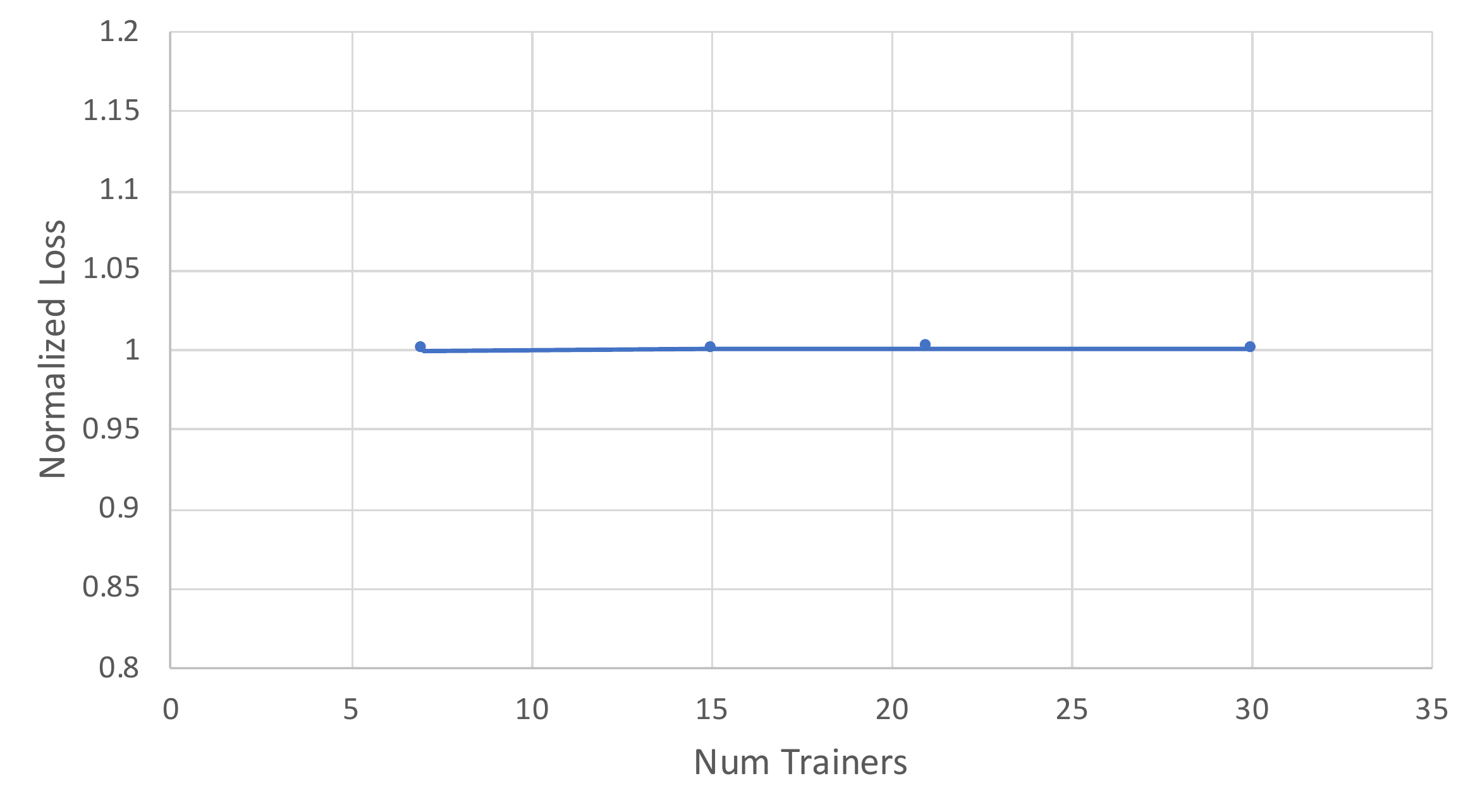}
  \caption{Quality as the number of trainers are increased.}
  \label{fig:training_quality}
\end{figure}

{\bf Asynchronous training} is well suited for a disaggregated design with use of dedicated parameter servers and trainers, that are relying on point-to-point \texttt{send}/\texttt{recv} communication or Remote Procedure Calls (RPC)~\cite{brock2019rdma}.

The compute intensive MLPs are replicated on different training processes and perform local weights updates based on the data samples they receive individually, only occasionally synchronizing with a master copy of the weights stored on the parameter server.  

The embedding tables are not replicated due to memory constraints, but are assigned to different training processes that receive asynchronous updates throughout training. Notice that because we use indices to access only a few of the embedding vectors in the forward pass, the simultaneous updates to an embedding table in the backward pass only collide when the indices used in the sparse lookups overlap between them.

It should be noted that both or a combination of these training algorithms can be used across different platforms such as a single node with multiple devices, multiple nodes or disaggregated setup of multiple trainers and parameter servers.

\begin{figure*}[t]
  \center 
  \includegraphics[width=0.95\columnwidth]{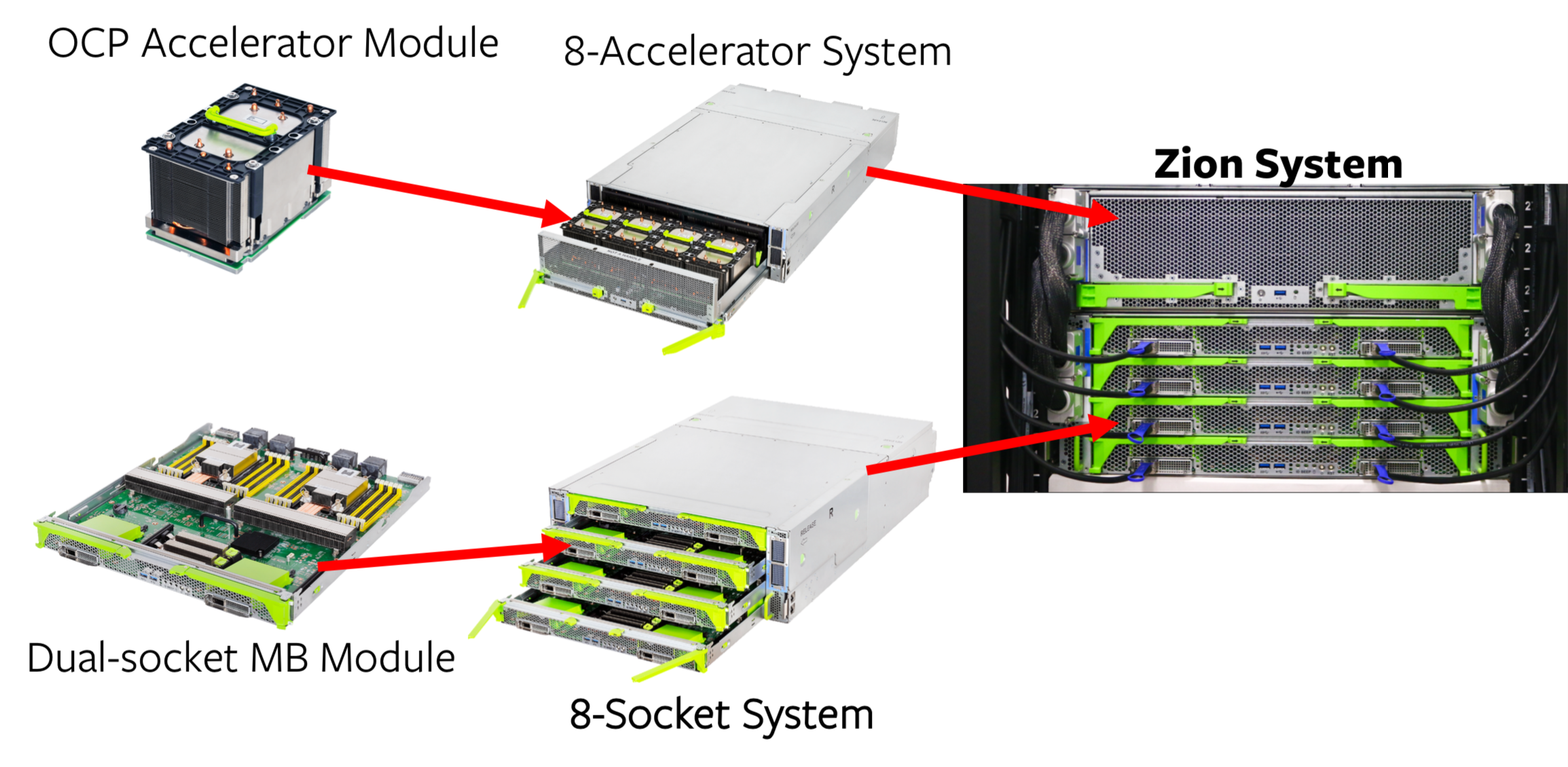}
  \hspace{0.1in}
  \includegraphics[width=0.95\columnwidth]{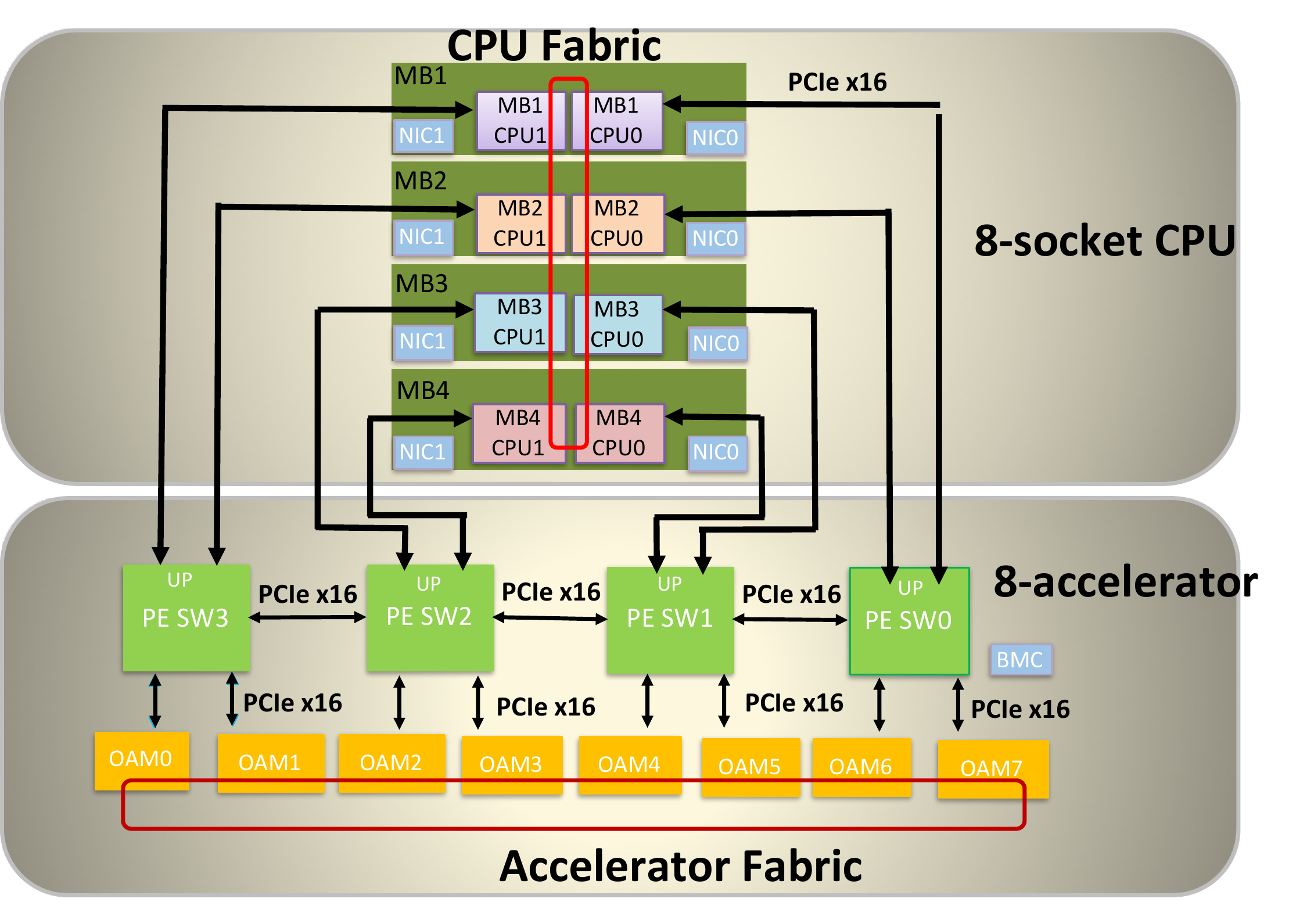}
  \centerline{\small{(a)} \hspace{3in} \small{(b)}  }
  \caption{(a) High-level overview of Zion system integration and (b) detailed block diagram of the Zion platform}
  \label{fig:zion_figure}
\end{figure*}

\subsection{Scalability \& Challenges}

The throughput of model training is important for fast prototyping and iterating on new ideas during model development. A single machine is not able to provide the throughput we need for our large recommendation models, and therefore we are heavily investing in scaling distributed training. 

Figure \ref{fig:trainer} shows the training throughput of one of our recent DLRMs. Here we use asynchronous training to avoid being bottlenecked by slow machines and/or interconnects. Also, we make sure the synchronization of the dense parameters is frequent enough, so that models will not diverge on different machines. As a result, we observe that the training throughput scales almost linearly with the number of hosts we use in one job, while model quality remains in an acceptable range, as is shown in Figure \ref{fig:training_quality}. 

In our experience, asynchronous training works well when a limited number of trainers is used, but with increasing number of trainers, we must incorporate synchronous training as an option into our system and hardware platforms.

\subsection{Interconnection Networks}

Aside from the standard \texttt{send/recv} communication primitives, the data- and model-parallelism used in distributed training give rise to two types of communication patterns: (i) \texttt{allreduce} operation that is often used to aggregate local parameter updates in the backward pass and (ii) \texttt{alltoall} operation that can be used to exchange the activations and error gradients across multiple nodes with different model weights.

The efficient synchronous or asynchronous implementation of these primitives relies on the support from the router or switch microarchitecture and the underlying interconnect network. As technology evolved through the years, the interconnection network fabric has varied. For example, the high performance computing (HPC) interconnect commonly used in the late 80s or 90s were often based on the low-radix topologies, such as 2D or 3D mesh or torus network~\cite{dally_kary_ncube, agarwal_icn}. As the pin bandwidth has increased it has been shown that the full bandwidth can be effectively utilized by partitioning it across an increasing number of ports, resulting in high-radix topologies~\cite{dragonfly} that were designed to reduce network diameter and cost~\cite{JKim05}. Also, it is important to note that HPC community has often relied on custom fabric and protocols, while in the data centers, in order to drive down the costs, the commodity interconnects are much more prevalent~\cite{AlFares08}.

\section{Zion Scale-up Training}
\label{sec:zion}

\subsection{Overview}

The building blocks of the Zion system consist of CPUs, accelerators and a flexible fabric that provides high performance while interconnecting these components~\cite{ZionHotChips}. Specifically,    Zion    decouples    memory,    compute,    and network  components  of  the  system,  allowing  each  to  scale independently as shown in Figure~\ref{fig:zion_figure}.  The  baseline Zion system  provides  8$\times$  NUMA CPU sockets with a large pool of DDR memory for capacity and  8$\times$  accelerators  to  provide the high   compute  capacity and also high bandwidth memory.

One of the challenges with leveraging accelerators for a training platform is determining which ones to use, given a large number of them that are becoming available.  It is infeasible to develop and enable a unique system for each one of the different accelerators. As a result, Facebook-led Open Accelerator Infrastructure (OAI) initiative ~\cite{oaiwiki} proposed to define vendor-agnostic common accelerator infrastructure, including a standard accelerator form factor Open Accelerator Module (OAM) that has been open sourced to the hardware  community. The OAM form factor abstracts the  various requirements  to  make  it  solution-agnostic  and  defined  as a common  form  factor   that  can  be  adopted  by  different accelerator  vendors.  The  common  form  factor along with the  baseboard enables using   multiple   accelerator   alternatives   with   the   same system  design.  

\begin{table}[t]
\center
\caption{Zion device comparisons.}
\begin{tabular}{l | l | l | l}
\multicolumn{2}{l|}{}  & CPU & Accelerator  \\
\hline
\hline
\multicolumn{2}{l|}{\# of devices}  & $\phantom{\sim}$8 & $\phantom{\sim}$8 \\
\hline
Compute in FP32 (TFlops) & aggregate & $\sim$20 & $\sim$100   \\
\hline 
Compute in FP16/BF16 (TFlops) & aggregate & $\sim$50 & $\sim$1000\\
\hline
Memory Capacity (TB) & aggregate & $\sim$2 & $\sim$0.2 \\
\hline
Memory Bandwidth (TB/s) & aggregate & $\sim$1 & $\sim$8 \\
\hline
Power (Watts) & per device & $\sim$100 & $\sim$200 \\
\hline
\end{tabular}
\label{tab:zion_devices}
\end{table}

The comparison between the characteristics of typical CPUs and accelerators is shown in Table~\ref{tab:zion_devices}. Notice that while the number of CPUs and accelerators in the system is the same, their compute and memory capabilities differ significantly. For example, the accelerator provides one or even two orders of magnitude higher compute (i.e., TFlops) as well as almost an order of magnitude higher memory bandwidth, all of which come at the cost of higher power. Also, accelerators rely on high-bandwith memory (HBM) that does not necessarily provide the same capacity as the DDR memory used in the CPUs -- thus, the CPUs provide an order of magnitude higher memory capacity. 

\begin{figure}[t]
 \center 
 \includegraphics[width=0.95\columnwidth]{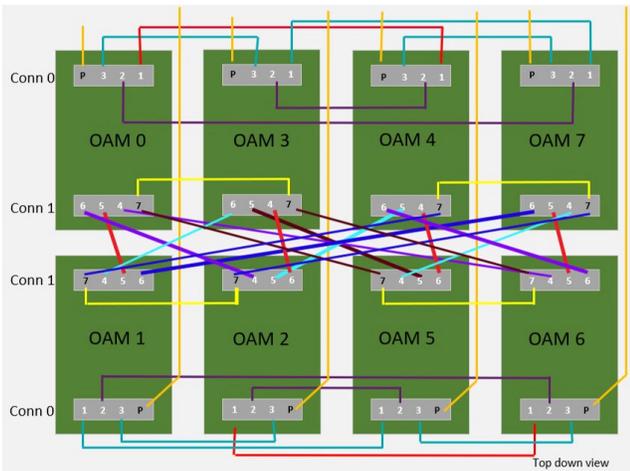}
 \caption{Accelerator fabric interconnecting layout in Zion.}
 \label{fig:zion_layout}
\end{figure}

\subsection{Accelerator Fabric Topology}

Another big  challenge  in  designing a common multi-source accelerator platform is  the  interconnect,  because  different vendors have  distinct  solutions  utilizing various topologies, protocols, and  number of  lanes/links. The Zion platform consists of 3 different types of interconnect fabrics -- the CPU fabric, the accelerator fabric and the PCIe interconnect that provides connectivity between CPUs and accelerators, see Figure~\ref{fig:zion_figure}. The CPU fabric options are limited by the vendors, but both PCIe and the accelerator interconnect are flexible, allowing for co-design to meet application needs.

The main components of any interconnection network include the topology, routing, flow control, and router micro-architecture~\cite{dallybook}.  Table~\ref{tab:icn_fabric} summarizes how these components differ from conventional high-performance and accelerator fabric interconnection networks. Overall the objectives of the two fabrics are slightly different -- HPC systems require low latency (and low network diameter) but also high bisection bandwidth.  In comparison, accelerator fabrics are less latency sensitive but often require high node-to-node bandwidth to efficiently support collective communication. 

The router micro-architecture impacts the topology and other components of the network. In HPC systems, given a topology, the routing algorithm, especially adaptive routing, is critical to exploit path diversity and maximize performance; however, most accelerator fabrics do not have hardware routers and therefore routing algorithms are not as critical because of the deterministic communication pattern.

In accelerator fabric without hardware support for communication, it often resembles ``store \& forward'' flow control, compared to ``cut-through'' flow control, since data is copied from one node to its neighboring node before being transmitted downstream. All nodes perform the same flow control at the same time, which results in high utilization across all nodes.

\begin{figure}[t]
 \center 
 \includegraphics[width=0.95\columnwidth]{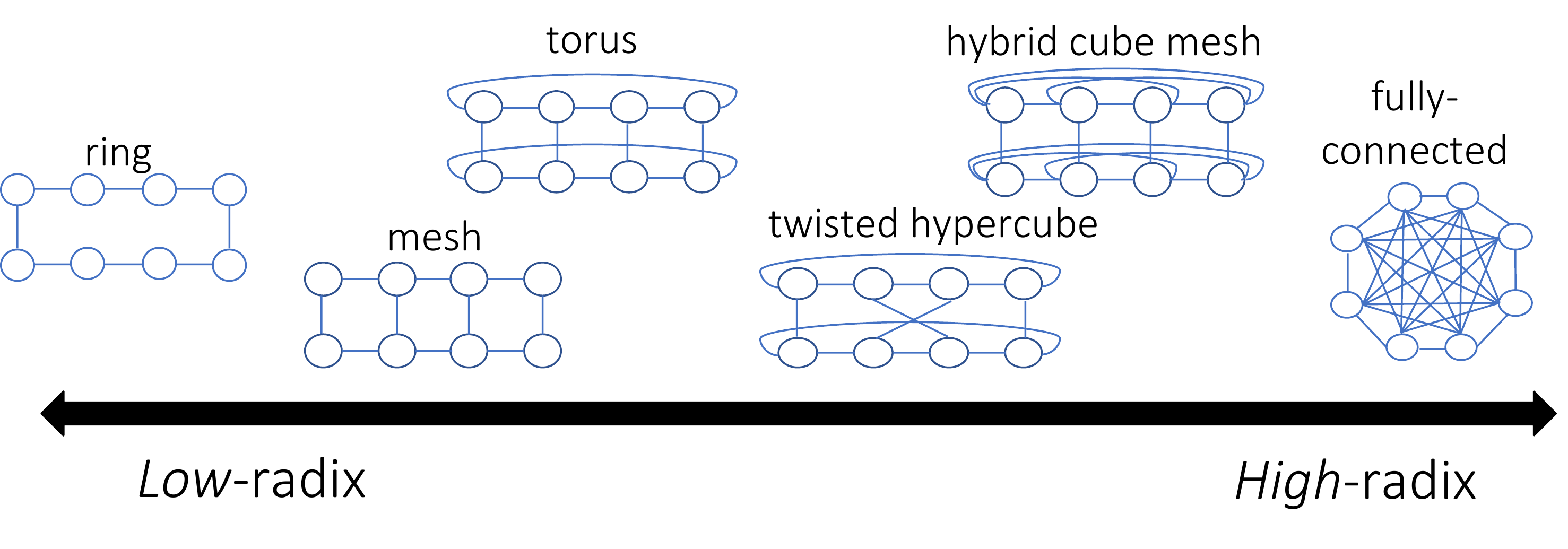}
 \caption{Different topology design space for an 8-node system.}
 \label{fig:8node}
\end{figure}

\begin{table}[t]
\caption{High-level comparison of interconnection networks.}
\begin{tabular}{l | l | l}
 & High Performance Interconnect & Accelerator Fabric \\
\hline
\hline
Topology & low-diameter, high    & high    \\
         & (bisection) bandwidth & (node) bandwidth  \\

\hline
Routing & adaptive routing  & deterministic routing \\
\hline
Flow Control & cut-through  & store \& forward \\
\hline
Fabric Design & router centric & node centric \\
\hline
\end{tabular}
\label{tab:icn_fabric}
\end{table}

In addition to software, there are many other physical design constraints for vendor agnostic topology design because each offering differs on many aspects and there is no established standard. An example of one such constraint is the accelerator interconnect link mapping, for some offering this may fan out on both sides and in other cases can be only on one side of the accelerator chip. The baseboard design presents routing length and other challenges to deal with the interconnect link insertion loss and also PCB layers/cost considerations. Figure~\ref{fig:zion_layout} is an illustration of top-down view of the OAM layout in Zion showing the connections between each OAM, with each colored link representing a different routing layer. As per the OAM specification each module has two mezzanine connectors (Conn $0/1$) at the south and north of each module and $1\times 16$ PCIe link to host (port $P$). Further each module has $7 \times 16$ interconnect links, shown in Figure~\ref{fig:zion_layout} as ports number from $1$ through $7$, among these $1-6$ are $\times 16$ links each and $7$ is split into $2 \times 8$ links to accommodate different topologies shown in Figure~\ref{fig:8node}.   

\subsection{Communication Pattern}

The Zion platform is designed to support both asynchronous and synchronous training algorithms, including support for model- and data-parallelism.  To support both types of parallelism we rely on the optimized \texttt{allreduce} and \texttt{alltoall}  communication  primitives implemented over the accelerator fabric topology.

\begin{figure}[t]
 \center 
 \includegraphics[width=0.84\columnwidth]{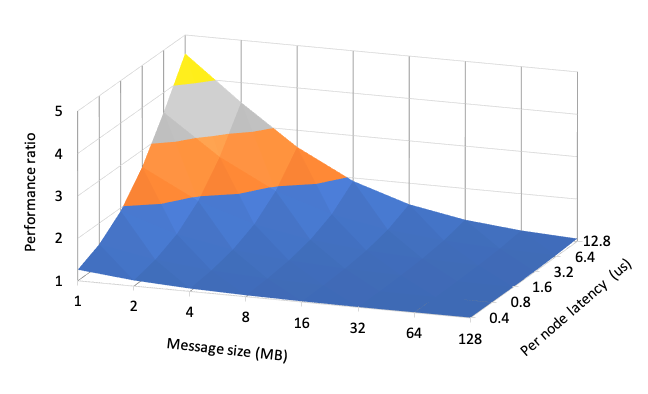}
 \centerline{\small{(a)}}
 \includegraphics[width=0.84\columnwidth]{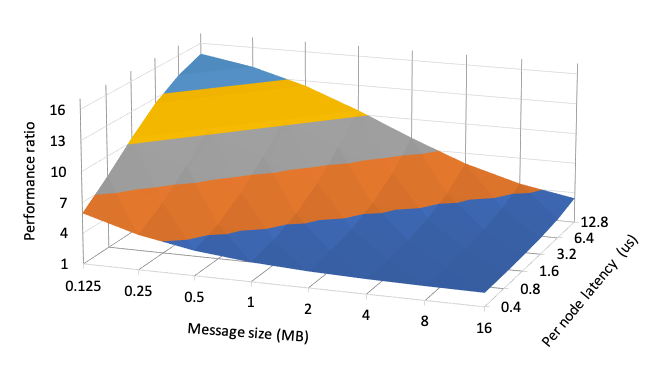}
 \vspace{-10pt}
 \centerline{\small{(b)}}
 \caption{Analytical performance model comparing ring vs fully connected topology for (a) \texttt{allreduce} and (b) \texttt{alltoall} communication primitive.}
 \vspace{-15pt}
 \label{fig:zion_comm_comparison}
\end{figure}

To understand the impact of accelerator topology across these communication primitives, we provide an analytical model comparing the added latency of going through a node for a low-radix ring topology and a high-radix fully-connected (FC) topology  for an 8-node system shown in Figure~\ref{fig:zion_comm_comparison}. We plot the performance ratio between the execution time predicted by the model for ring and FC topology, ignoring additional software overheads often found in practice \cite{AngLi2019}. In our model we fix per node bandwidth $B$, and vary the message size $M$ and per-node latency $\alpha$, to simulate the traffic pattern.  The execution time for \texttt{allreduce} ($T\_AR$) for the two topologies, using a ring algorithm,  can be summarized as 

\begin{eqnarray}
 T\_AR_{ring} & = &  2 \frac{M}{B} \frac{p - 1}{p}  + 2 \alpha (p - 1)  \\
 T\_AR_{FC} & = &  2 \frac{M}{b} \frac{1}{p}  + 2 \alpha
\end{eqnarray}

\noindent
where $p$ is the number of nodes and $b$ is the bandwidth per channel. In the comparison, $p=8$ and we assume total bandwidth from each accelerator node is  constant -- thus, the ring has ``fatter'' channels while the fully-connected topology has ``thinner'' channels. Thus, the bandwidth per channel in the FC is effectively $b = B / (p-1)$.  By plugging this value into $T\_AR_{FC}$, the bandwidth component of performance across the two topologies can be seen to be identical. 

Figure~\ref{fig:zion_comm_comparison}(a) plots ratio $T\_AR_{ring} / T\_AR_{FC}$. Note that the ratio higher than 1 represents region where FC outperforms the ring topology. As the message size increases, the bandwidth component dominates and therefore the physical topology has no impact.  However, as the message size decreases and per-node latency increases, the latency component dominates and therefore the performance benefit from the FC increases significantly (it minimizes the network hop count). 

The execution time for \texttt{alltoall} ($T\_A2A$) for the two topologies, using a ring algorithm,  can be summarized as 

\begin{eqnarray}
 T\_A2A_{ring} &  = &  \gamma ( \frac{M}{B}\frac{1}{p} + \alpha) \\
 T\_A2A_{FC} & = &  \frac{M}{B} \frac{p-1}{p}  + \alpha
\end{eqnarray}

\noindent
where $\gamma = 2 \sum_{h=1}^{q} h + r \frac{p}{2} $, with quotient $q = (p-1) // 2$, remainder $r = (p-1) \% 2$ and \# of hops between nodes $h$.

For FC topology, the cost of $T\_A2A$ is approximately 1/2 of $T\_AR$ because \texttt{alltoall} communication only consists of a single phase, while \texttt{allreduce} based on the ring algorithm consists of 2 phases.  In comparison, for the ring topology, \texttt{alltoall} requires each node to send messages to all other nodes and performance is proportional to the sum of the hop count between each pair of nodes $\gamma$, while for FC the network diameter is 1.  For the ring topology, we assume \texttt{alltoall} communication is done in multiple steps where each node sends message to destination that is 1-hop away, 2-hop away, etc and ensure all channels are fully utilized. 
Thus, for large message sizes, the benefit from FC approaches 
$\gamma / (p-1)$ or $\sim$2.3 in the 8-node example.  For smaller message size, the benefit from FC is much higher because of the latency component as shown in Figure~\ref{fig:zion_comm_comparison}(b). 

In distributed synchronous training of DLRMs the message size for \texttt{allreduce} is often around $10$MB -- thus, the physical topology does not have significant impact on overall communication cost. However, for \texttt{alltoall} communication the message size is much smaller (e.g., on the order of $100$KB), therefore FC topology can provide up to 3$\times$ improvement in communication cost compared to the ring topology.  This pushes distributed DLRM training characteristics towards HPC workloads that rely on global traffic patterns
and where low-diameter topology can provide significant benefits. 

\subsection{Hardware Design Flexibility}
\begin{figure}[t]
 \center 
 \includegraphics[width=0.45\columnwidth]{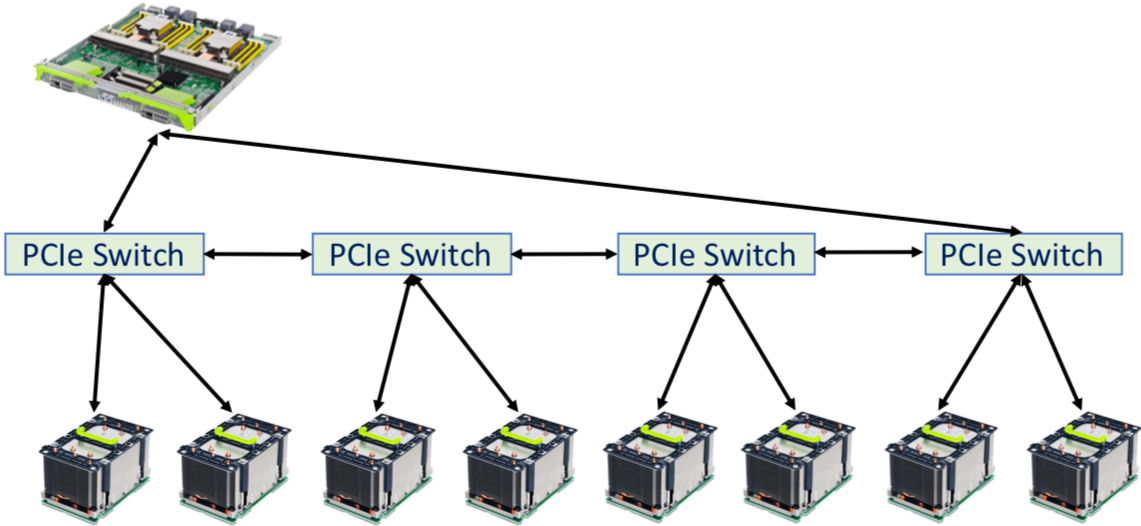}
 \includegraphics[width=0.45\columnwidth]{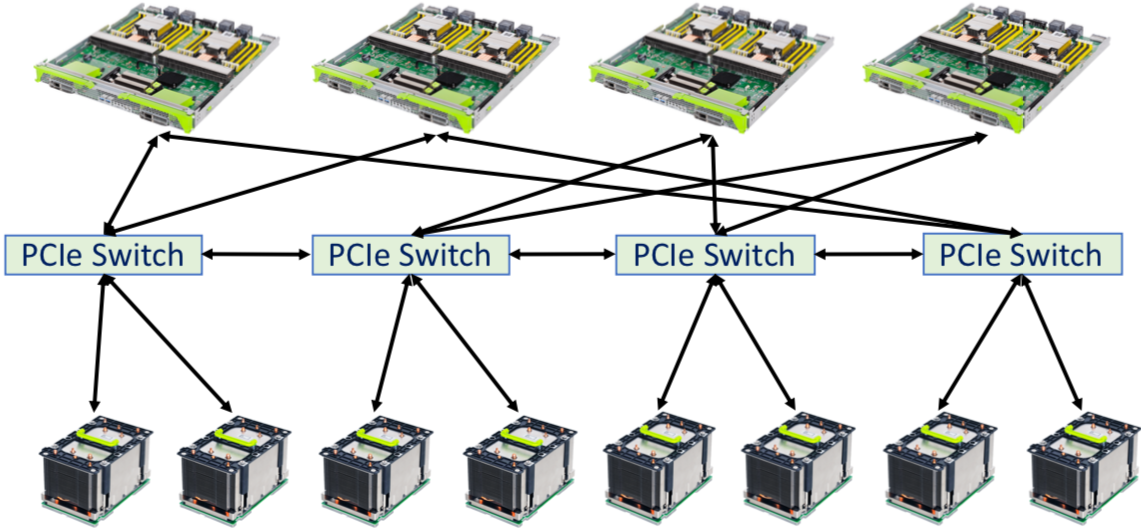}
 \centerline{\small{(a)} \hspace{2in} \small{(b)}}
 \caption{System flexibility by leveraging only (a) single 2-socket CPU system and (b) four 2-socket CPUs within the Zion platform.  Even with a single 2-socket system, the interconnect of the PCIe switches enable full connectivity between the CPUs and the accelerators. }
 \label{fig:zion_flexibility}
 \vspace{-10pt}
\end{figure}

Although Zion hardware is designed mainly for Facebook deep learning recommendation workload, the design is formed by 4 identical dual-socket motherboard modules and is flexible enough to be configured for example to use 2 sockets plus 8 accelerators, as shown on Figure~\ref{fig:zion_flexibility} (a). This 2 CPU sockets configuration may be used for other workloads which do not require large memory footprint, such as computer vision and natural language processing. On the other hand, Figure~\ref{fig:zion_flexibility} (b) shows the standard 8 socket plus 8 accelerators configuration.

\section{Discussion on Scale-out Training}

The Zion system described earlier provides significant amount of compute and memory capacity to support large neural network models.  However, one limitation of its memory system is that aggregate $1.5$TB of DDR memory does not provide high enough memory bandwidth. Moreover, it is anticipated that model complexity and the amount of input data will continue to grow as shown on Figure~\ref{fig:model_size}, resulting in the need for increased training throughput. 

The increasing demand can be best addressed by scaling out training to multiple nodes and increasingly leveraging accelerators. This translates into training system beyond a scale-up solution of single super-node (e.g. Zion)  to scale-out systems with multiple such supernodes. However, distributed  training introduces additional fabric and connectivity requirements across multiple nodes to efficiently support asynchronous and synchronous training with data- and model-parallelism. In this section, we discuss three vectors, namely network topology, network transport, and implementation of optimized collective primitives, that have significant implications on the design of scale-out systems for synchronous training of DLRMs. 

\begin{table}[t]
 \centering
 \caption{Description of different scale-out training systems. }
 \begin{tabular}{ |c|c|c| }\hline
 \diagbox[width=8em]{Protocol}{Topology }
             & \makebox{Flat} & \makebox{Hierarchical}\\\hline 
 Homogeneous &  \begin{tabular}{@{}c@{}} TPU~\cite{TPUhotchips} \\  
 Habana~\cite{GaudiHotChips} \end{tabular} & \begin{tabular}{@{}c@{}} Habana HLS-1~\cite{GaudiHotChips} \\ Intel SpringCrest~\cite{IntelSpringCrest}  \end{tabular}  \\ \hline
 Heterogeneous  & N/A  & \begin{tabular}{@{}c@{}}Zion~\cite{ZionHotChips} \\ DGX SuperPod~\cite{dgxpod} \end{tabular}  \\ \hline
 \end{tabular}
 \label{tab:scaleout_systems}
\end{table}

\subsection{Network Topology}

Different scale-out training systems have been proposed, developed and successfully deployed by the industry as summarized in Table~\ref{tab:scaleout_systems}. These systems can  be classified as flat or hierarchical based on the fabric organization -- flat topology uses a single global topology (e.g., 2D or 3D torus used in the TPU system~\cite{TPUhotchips}) while a hierarchical organization consists of a ``supernode'' with several accelerators, which is used as the building block to scale-out to larger number of nodes (e.g. DGX SuperPod ~\cite{dgxpod}). 

Another  difference in scale-out fabric is whether the fabric is accelerator- or CPU-centric.  We define an accelerator fabric similar to the definition in the Zion system -- if the accelerators are used as the building block to scale-out, the system can be referred to as using accelerator-centric fabric, while if the CPU is used to scale-out, the system is CPU-centric. While the Zion system itself uses an accelerator-centric fabric, the Zion system when scaling-out is heterogeneous and CPU-centric since the 100 GbE network interface of the CPUs are used to scale-out. 

The scale-out systems can further be categorized based on the communication protocol -- a homogeneous system leverages the same link/interconnect technology for the entire system while a heterogeneous system exploits different interconnect technologies. We note that while different interconnect protocols (e.g., Infiniband~\cite{Shanley2002}, RoCE~\cite{mittal2018revisiting}, NVlink~\cite{Foley2017}, etc.) are available, heterogeneous systems can be bottlenecked by the channels that provide the smallest amount of bandwidth.

\begin{figure}[t]
 \center 
 \includegraphics[width=0.95\columnwidth]{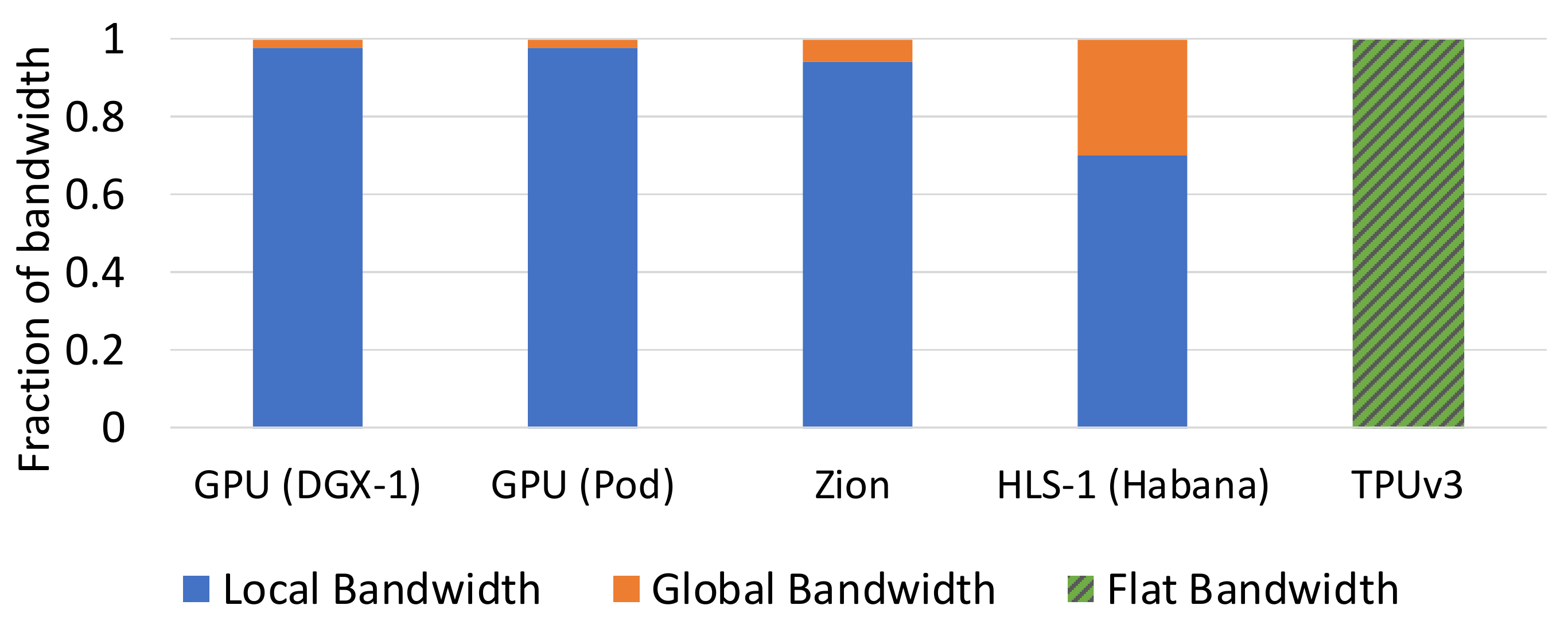}
 \caption{Comparison of local vs global bandwidth for different scale-out systems.  The TPU system does not differentiate between local and global bandwidth since it is flat topology.}
 \label{fig:localvsglobal_bandwidth}
\end{figure}

\begin{figure}[t]
 \center 
 \includegraphics[width=0.95\columnwidth]{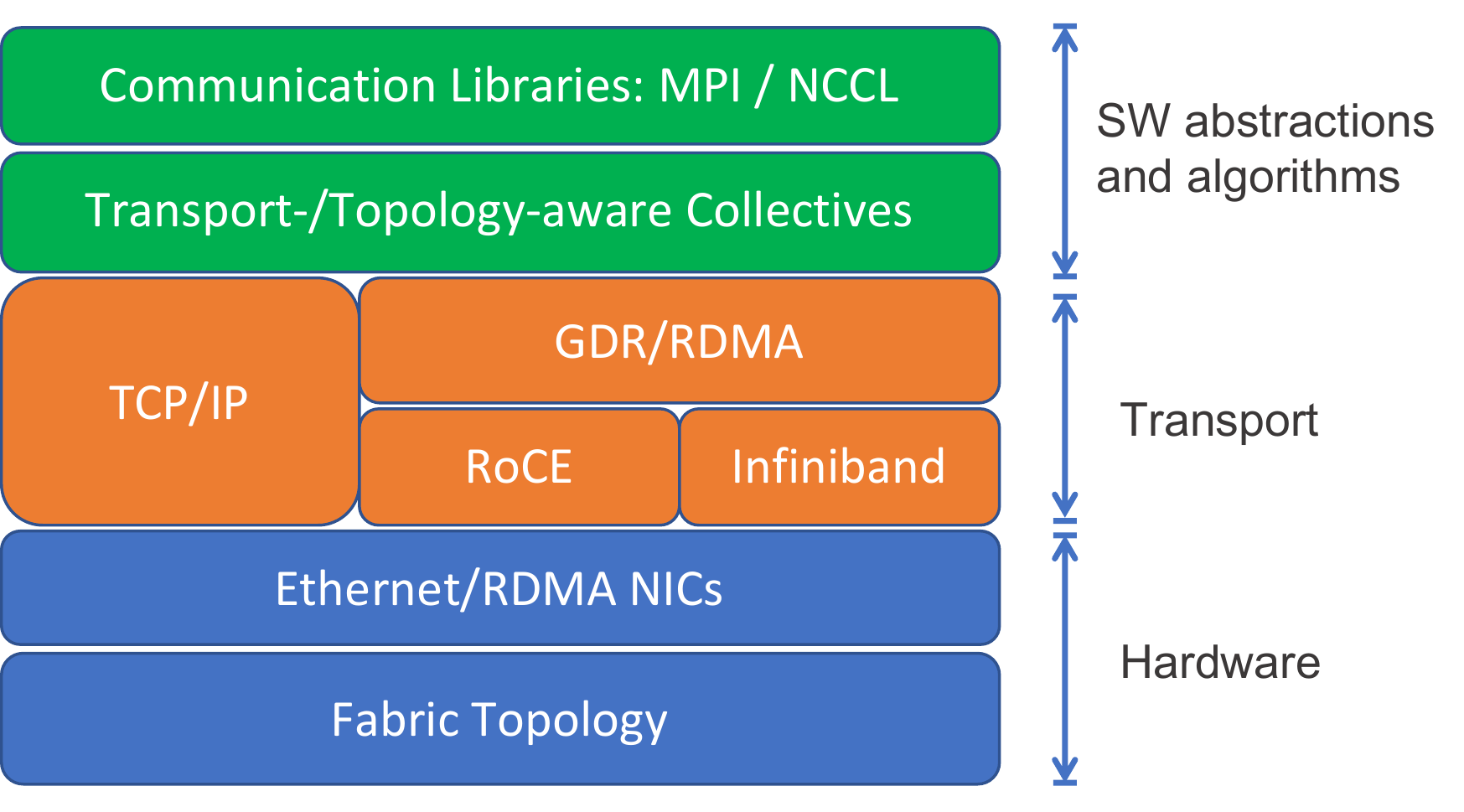}
 \caption{A view of interconnect software and hardware stack}
 \label{fig:rdma_rpc_stack}
 \vspace{-5pt}
\end{figure}

Also, we delineate a distinction between \emph{global} and \emph{local} bandwidth in the scale-out systems. Local bandwidth refers to the bandwidth for communicating between physically neighboring nodes, while global bandwidth is used to provide connectivity across the (global) system. In Figure~\ref{fig:localvsglobal_bandwidth}, we compare the ratio of local vs global bandwidth across different scale-out systems.  For the NVidia GPU systems, the amount of local bandwidth dominates overall system bandwidth as each of the GPU nodes has 300 GB/s of bandwidth (through NVLinks) for local communication while the scale-out bandwidth through the Infiniband  is 4x100 Gbps and shared across 8 GPUs.  The amount of scale-out bandwidth increases for the GPU(Pod) but it is shared among larger number of GPUs and thus, the ratio is similar.  For the Zion system, the fraction of global bandwidth is slightly higher since there are 8 CPU network interfaces to provide the scale-out bandwidth.  In comparison, the HLS-1 system from Habana has 10 RoCE channels from the Gaudi accelerator and uses 7 channels for intra-box connectivity while 3 channels are used for scale-out or global bandwidth. In contrast,  the flat topology (i.e. TPU) does not differentiate between local and global bandwidth since the fabric is organized as a single, global topology.  

\subsection{Network Transport}

\begin{figure*}[t]
 \center 
 \includegraphics[width=1.5\columnwidth]{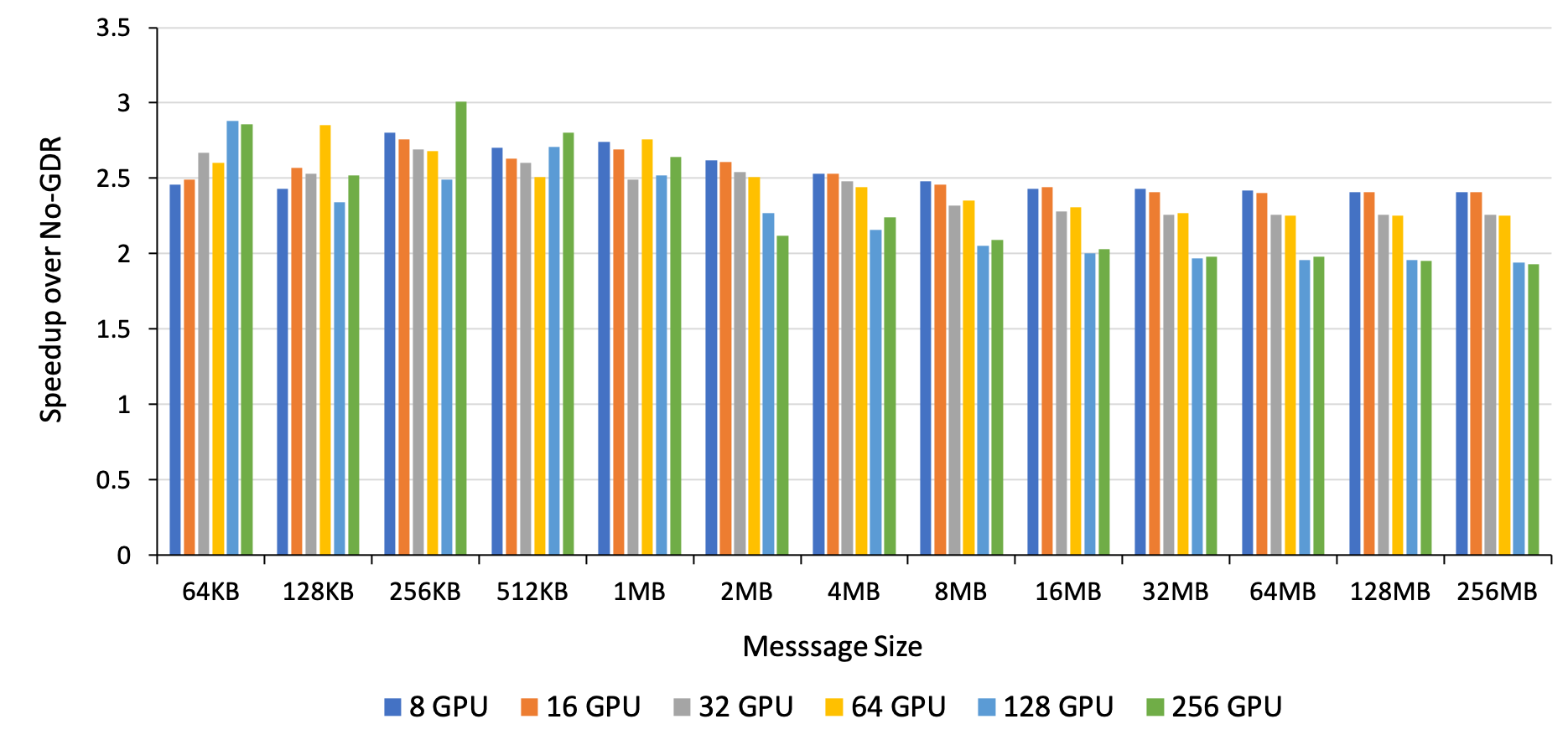}
 \caption{Speedup of GDR over Non-GDR \texttt{allreduce} scale-out communication}
 \label{fig:gdr-scaleout-speedup}
 \vspace{-10pt}
\end{figure*}

The network transport layer is key in implementing communication primitives for a given fabric topology. Hyper-scale data-centers have typically used kernel-based TCP/IP and Ethernet as the network transport for decades. However, accelerator-centric fabric, efficient scale-out communication entails moving data directly from one accelerator to another without host CPU involvement - requiring a transport that supports Remote Direct Memory Access (RDMA) or similar technology between accelerators. 

Infiniband and RDMA over Converged Ethernet (RoCE) are two of the most popular examples of RDMA transport in use today, see Figure~\ref{fig:rdma_rpc_stack}. These protocols are supported by RDMA Network Interface Controllers (RNICs), which are PCIe devices that are comprised of a sophisticated DMA and transport engines, and include various management capabilities. RNICs use Verbs interface, which provides operations for resource management and data transfer, including send/receive and RDMA read/write operations. Today’s RNICs typically offer the following capabilities: 

\begin{itemize}
    \item \emph{Zero} copy -- direct data exchange between buffers 
    \item \emph{Transport offload} -- segment/assembly, retransmission 
    \item \emph{Kernel bypass} -- RNICs hardware access from user mode 
    \item \emph{Flow control} -- credit-based control, end-to-end reliability 
\end{itemize}
Further, support for RDMA and Verbs by itself is not sufficient for enabling optimized network transports for accelerator-centric fabric. For instance, NVidia introduced a series of enhancements to enable RNICs to directly access GPU memory, culminating in GPUDirect RDMA (GDR)~\cite{GDR2019}. GDR provides direct path for data movement between GPU memory and Infiniband/RoCE NICs. It eliminates significant overheads resulting from PCIe over-subscription\footnote{On GPU systems, such as NVidia’s DGX-1, one 1x16 PCIe link to the host is shared by three devices each driving 1x16 PCIe lanes. This PCIe over-subscription on the host PCIe link results in significant overheads when moving data to/from system memory.} and additional copy through the host/system memory.   

To illustrate the performance advantages of using GDR, we evaluate GDR on Infiniband fabric against a highly optimized Non-GDR baseline, that requires one extra copy but also uses RDMA transport over lossless Infiniband fabric\footnote{We point out that alternate host kernel based transports, such as TCP/IP over Ethernet incur two copies.}.  We run the benchmarks starting with one DGX-1 node (8 GPUs) all the way up to $32$ nodes ($256$ GPUs). We force all communication via Infiniband to better isolate the benefits of GDR since it primarily affects the scale-out data path. We collect the performance data for $64$KB to $256$MB message sizes, representing typical message sizes found in machine learning workloads. 

In this scenario Figure \ref{fig:gdr-scaleout-speedup} presents the speedup of GDR over Non-GDR for NCCL \texttt{allreduce} collective\footnote{NCCL \texttt{all2all} collective is not yet available}. Note that given that NICs and GPUs are on the same PCIe switch with a single link to the CPU, in the non-GDR case send/recv flows go through the PCIe link to the CPU causing the bandwidth to be halved. GDR avoids this problem attaining ~$2\times$ higher bandwidth and corresponding speedup for large messages. On the other hand, for small message sizes, GDR also has an advantage in terms of latency because it avoids intermediate copy to host/system memory, which explains why we see an even greater speedup in this case. 

Thus RDMA  transports (e.g. GDR) overcome fundamental limitations in host-accelerator connectivity and are important for enabling efficient scaling on accelerator-centric fabric. They are also often used to build higher level abstractions that may be exposed in DL frameworks~\cite{tensorflow2015-whitepaper,paszke2017automatic,chen2015mxnet,jia2014caffe}. 

\subsection{Transport- and Topology-aware Collectives}

Recall that to support both asynchronous and synchronous training with data- and model-parallelism we rely on the \texttt{allreduce} and \texttt{alltoall}  communication  primitives. These primitives have varying characteristics depending on size of data being communicated, the number and topology of nodes involved in the communication. 

For both \texttt{allreduce} and \texttt{alltoall}, the data-sizes of interest vary from a few KBs to several MBs. Typically the smaller data sizes are more sensitive to latency and as the data size increases bandwidth becomes more important. Also, unlike the collective communication requirements from \texttt{allreduce}, in the \texttt{alltoall} all nodes need to communicate with all other nodes and thus, the average hop count of the topology has a significant impact on overall performance. 

In short, collective implementations need to be aware of (a) network transport (GDR/RDMA vs. TCP), and (b) network topology. For instance, Message Passing Interface (MPI) and NVIDIA Collective Communication Library (NCCL), implement topology- and transport-aware collectives today.  

Future scale-out systems need to be co-designed taking into account all of these considerations, in particular including fabric topology, transport, and optimized collectives.

\section{Related work}

Given the growing importance of distributed training of deep learning models, there has been a number of papers on workload characterization and scale-out solutions~\cite{Chilimbi2014,Gibiansky2017,Mayer2019}, including Alibaba's Platform of Artificial Intelligence (PAI)~\cite{Alibaba-PAI} and EFLOPS~\cite{Alibaba-EFLOPS}, which detail algorithm and system co-design for high performance distributed training platforms. Other works such as DaDianNao~\cite{DaDianNao} and ScaleDeep~\cite{ScaleDeep}, also showcase similar results. We provide an overview of some of the well established and more recent hardware platforms that have tried to address the corresponding challenges, with emphasis on their scale-out approach and communication.

The NVidia GPUs have been commonly used for deep learning training, and are often organized into the DGX appliance. DGX-1 consists of 8 GPUs that are interconnected with high-bandwidth NVlink. DGX-2 was announced and provides 16 GPUs that are interconnected not only with NVlink for high-bandwidth but also NVSwitch to provide connectivity between all 16-GPUs. Further, Infiniband may be used to scale beyond a single DGX-1 or DGX-2 system. We point out that NVidia DGX Pod is based on DGX-1 system.
  
The Google TPU Pod~\cite{TPUhotchips, TPU_ISCA} (v2 and v3) is an example of a DNN supercomputer that is tightly integrated with custom accelerator (TPU), high-bandwidth memory and a custom interconnect to enable a supercomputing pod. The TPUs are interconnected together with a high-speed link that supports a custom protocol, instead of commonly available communication protocol, to reduce communication overhead and an integrated router is supported in each TPU~\cite{TPUtutorial}. While the system is organized hierarchically with a board consisting of 4 TPUs and a rack consisting of multiple boards, the global topology to scale-out is a low-radix 2D toridal mesh.

The Habana Gaudi training processors~\cite{GaudiHotChips} provide integrated compute and networking by supporting RoCE within the training processor. Each Gaudi has 10 ports of 100Gb Ethernet and each port can be used for either internal or external (scale-out) connectivity. The HLS-1 system with 8 Gaudi leverages 7 ports for fully connected topology within the server while the remaining 3 ports are used to scale-out.
 
On the other hand, Cerebras has proposed wafer-scale training system~\cite{WaferScale} and built one of the largest chips ever created with over 1 trillion transistors. By integrating all the cores in a single wafer,  it enables over 100 Pbit/s of fabric bandwidth and 8 PByte/s of memory bandwidth.  The cores (or ``chips'')  within the wafer are interconnected through a 2D mesh topology. While wafer-scale provides significant compute and bandwidth (both memory and interconnect), it presents some unique challenges from wafer-scale computing -- including yield, power/thermal challenges, as well as packaging constraints.

Huawei has also recently announced its Ascend 910~\cite{HuaweiTraining} training solution. While some of the details are not clear, 8 Ascend 910 chips are integrated into a single ``AI server,'' similar to a NVidia DGX-1, with a proprietary high-speed link technology (HCCS) providing 720 Gbps of bandwidth for NUMA connections within the AI server. In addition, 200 Gbps of bandwidth is provided through RoCE interface for scale-out to create an Ascend cluster with 1K to 2K nodes.

Finally, Intel announced Spring Crest Deep Learning Accelerator for training~\cite{IntelSpringCrest} to enable scalable deep learning. Each Spring Crest provides 64 lanes of SerDes for a total aggregate bandwidth of 3.58 Tbps and each chip also has a fully programmable router that enables multiple ``glueless'' topologies -- i.e., a server consisting of 8 Spring Crest can be directly interconnected  to another server and can scale up to 1K nodes.

\section{Summary}

In this paper we have discussed different aspects of training of DLRMs. We have touched on asynchronous and synchronous training, including the mapping of MLPs and embeddings onto data- and model-parallel patterns of computation. Further, we have outlined how data- and model-parallelism patterns map to \texttt{allreduce} and \texttt{all2all} communication primitives, respectively. Also, we have shown how the performance of these primitives is influenced by fabric topology and interconnect design. 

We have described the design of the Zion scale-up system and how it can be used for training of DLRMs. We have also reviewed and compared the organization of some of the existing scale-out systems. Finally, we have outlined several important considerations for future scale-out systems, including fabric organization, network transport and topology-aware communication primitives.

\balance
\bibliographystyle{IEEEtranS}
\bibliography{refs}

\begin{thebibliography}{10}
\providecommand{\url}[1]{#1}
\csname url@samestyle\endcsname
\providecommand{\newblock}{\relax}
\providecommand{\bibinfo}[2]{#2}
\providecommand{\BIBentrySTDinterwordspacing}{\spaceskip=0pt\relax}
\providecommand{\BIBentryALTinterwordstretchfactor}{4}
\providecommand{\BIBentryALTinterwordspacing}{\spaceskip=\fontdimen2\font plus
\BIBentryALTinterwordstretchfactor\fontdimen3\font minus
  \fontdimen4\font\relax}
\providecommand{\BIBforeignlanguage}[2]{{%
\expandafter\ifx\csname l@#1\endcsname\relax
\typeout{** WARNING: IEEEtranS.bst: No hyphenation pattern has been}%
\typeout{** loaded for the language `#1'. Using the pattern for}%
\typeout{** the default language instead.}%
\else
\language=\csname l@#1\endcsname
\fi
#2}}
\providecommand{\BIBdecl}{\relax}
\BIBdecl

\bibitem{dgxpod}
\BIBentryALTinterwordspacing
``{NVidia DGX Pod},'' 2018. [Online]. Available:
  \url{https://www.nvidia.com/en-us/data-center/dgx-pod-reference-architecture/}
\BIBentrySTDinterwordspacing

\bibitem{GDR2019}
\BIBentryALTinterwordspacing
``{NVidia GPUDirect RDMA},'' 2019. [Online]. Available:
  \url{https://docs.nvidia.com/cuda/gpudirect-rdma/index.html}
\BIBentrySTDinterwordspacing

\bibitem{oaiwiki}
\BIBentryALTinterwordspacing
``{Open Accelerator Infrastructure (OAI)},'' 2019. [Online]. Available:
  \url{https://www.opencompute.org/wiki/Server/OAI}
\BIBentrySTDinterwordspacing

\bibitem{PaddlePaddle2019}
``Parallel distributed deep learning: Machine learning framework from
  industrial practice,'' 2019.

\bibitem{tensorflow2015-whitepaper}
M.~Abadi, A.~Agarwal, P.~Barham, E.~Brevdo, Z.~Chen, C.~Citro, G.~S. Corrado,
  A.~Davis, J.~Dean, M.~Devin, S.~Ghemawat, I.~Goodfellow, A.~Harp, G.~Irving,
  M.~Isard, Y.~Jia, R.~Jozefowicz, L.~Kaiser, M.~Kudlur, J.~Levenberg,
  D.~Man\'{e}, R.~Monga, S.~Moore, D.~Murray, C.~Olah, M.~Schuster, J.~Shlens,
  B.~Steiner, I.~Sutskever, K.~Talwar, P.~Tucker, V.~Vanhoucke, V.~Vasudevan,
  F.~Vi\'{e}gas, O.~Vinyals, P.~Warden, M.~Wattenberg, M.~Wicke, Y.~Yu, and
  X.~Zheng, ``{TensorFlow}: Large-scale machine learning on heterogeneous
  systems,'' 2015.

\bibitem{agarwal_icn}
A.~{Agarwal}, ``Limits on interconnection network performance,'' \emph{IEEE
  Transactions on Parallel and Distributed Systems}, vol.~2, no.~4, pp.
  398--412, 1991.

\bibitem{AlFares08}
M.~Al-Fares, A.~Loukissas, and A.~Vahdat, ``A scalable, commodity data center
  network architecture,'' in \emph{Proc. 32nd Int. Symp. Computer Architecture
  ({ISCA})}, 2005.

\bibitem{borisyuk2018rosetta}
F.~Borisyuk, A.~Gordo, and V.~Sivakumar, ``Rosetta: Large scale system for text
  detection and recognition in images,'' in \emph{Proc. 24th Int. Conf.
  Knowledge Discovery \& Data Mining ({KDD})}, 2018, pp. 71--79.

\bibitem{bottou2016optimization}
L.~Bottou, F.~E. Curtis, and J.~Nocedal, ``Optimization methods for large-scale
  machine learning,'' \emph{CoRR}, vol. 1606.04838, 2016.

\bibitem{brock2019rdma}
B.~Brock, Y.~Chen, J.~Yan, J.~D. Owens, A.~Buluç, and K.~Yelick, ``{RDMA} vs.
  {RPC} for implementing distributed data structures,'' \emph{CoRR}, vol.
  1910.02158, 2019.

\bibitem{TPUtutorial}
C.~Chao and B.~Saeta, ``Cloud {TPU}: Codesigning architecture and
  infrastructure,'' in \emph{Hot Chips 31 Symposium, Palo Alto, CA, USA}, 2019.

\bibitem{chen2015mxnet}
T.~Chen, M.~Li, Y.~Li, M.~Lin, N.~Wang, M.~Wang, T.~Xiao, B.~Xu, C.~Zhang, and
  Z.~Zhang, ``{MXNet}: A flexible and efficient machine learning library for
  heterogeneous distributed systems,'' \emph{CoRR}, vol. 1512.01274, 2015.

\bibitem{DaDianNao}
Y.~{Chen}, T.~{Luo}, S.~{Liu}, S.~{Zhang}, L.~{He}, J.~{Wang}, L.~{Li},
  T.~{Chen}, Z.~{Xu}, N.~{Sun}, and O.~{Temam}, ``{DaDianNao}: A
  machine-learning supercomputer,'' in \emph{Proc. 47th Annual IEEE/ACM Int.
  Symp. Microarchitecture ({MICRO})}, 2014, pp. 609--622.

\bibitem{Chilimbi2014}
T.~Chilimbi, Y.~Suzue, J.~Apacible, and K.~Kalyanaraman, ``Project {Adam}:
  Building an efficient and scalable deep learning training system,'' in
  \emph{Proc. 11th Symp. Operating Systems Design and Implementation ({OSDI})
  14)}, 2014, pp. 571--582.

\bibitem{covington2016deep}
P.~Covington, J.~Adams, and E.~Sargin, ``Deep neural networks for youtube
  recommendations,'' in \emph{Proc. 10th ACM Conf. Recommender Systems}, 2016,
  pp. 191--198.

\bibitem{dally_kary_ncube}
W.~J. {Dally}, ``Performance analysis of k-ary n-cube interconnection
  networks,'' \emph{IEEE Transactions on Computers}, vol.~39, no.~6, pp.
  775--785, 1990.

\bibitem{dallybook}
W.~Dally and B.~Towles, \emph{Principles and Practices of Interconnection
  Networks}.\hskip 1em plus 0.5em minus 0.4em\relax San Francisco, CA, USA:
  Morgan Kaufmann Publishers Inc., 2003.

\bibitem{das2016distributed}
D.~Das, S.~Avancha, D.~Mudigere, K.~Vaidynathan, S.~Sridharan, D.~Kalamkar,
  B.~Kaul, and P.~Dubey, ``Distributed deep learning using synchronous
  stochastic gradient descent,'' \emph{CoRR}, vol. 1602.06709, 2016.

\bibitem{Dean2012}
J.~Dean, G.~Corrado, R.~Monga, K.~Chen, M.~Devin, M.~Mao, M.~A. Ranzato,
  A.~Senior, P.~Tucker, K.~Yang, Q.~V. Le, and A.~Y. Ng, ``Large scale
  distributed deep networks,'' \emph{Advances in Neural Information Processing
  Systems ({NIPS}) 25}, pp. 1223--1231, 2012.

\bibitem{Alibaba-EFLOPS}
J.~Dong, Z.~Cao, T.~Zhang, J.~Ye, S.~Wang, F.~Feng, L.~Zhao, X.~Liu, L.~Song,
  L.~Peng, Y.~Guo, X.~Jiang, L.~Tang, Y.~Du, Y.~Zhang, P.~Pan, and Y.~Xie,
  ``{EFLOPS}: Algorithm and system co-design for a high performance distributed
  training platform,'' \emph{Proc. IEEE Int. Symp. High-Performance Computer
  Architecture (HPCA)}, 2020.

\bibitem{Foley2017}
D.~Foley and J.~Danskin, ``Ultra-performance {Pascal GPU and NVLink}
  interconnect,'' \emph{Proc. IEEE/ACM Int. Symposium on Microarchitecture
  ({MICRO})}, vol.~37, pp. 7--17, 2017.

\bibitem{Gibiansky2017}
A.~Gibiansky, ``Bringing {HPC} techniques to deep learning,'' \emph{Baidu
  technical blog}, 2017.

\bibitem{Gupta2019}
U.~Gupta, X.~Wang, M.~Naumov, C.~Wu, B.~Reagen, D.~Brooks, B.~Cottel, K.~M.
  Hazelwood, B.~Jia, H.~S. Lee, A.~Malevich, D.~Mudigere, M.~Smelyanskiy,
  L.~Xiong, and X.~Zhang, ``The architectural implications of facebook's
  dnn-based personalized recommendation,'' \emph{CoRR}, vol. 1906.03109, 2019.

\bibitem{hazelwood2018applied}
K.~Hazelwood, S.~Bird, D.~Brooks, S.~Chintala, U.~Diril, D.~Dzhulgakov,
  M.~Fawzy, B.~Jia, Y.~Jia, A.~Kalro, J.~Law, K.~Lee, J.~Lu, P.~Noordhuis,
  M.~Smelyanskiy, L.~Xiong, and X.~Wang, ``Applied machine learning at
  {Facebook}: A datacenter infrastructure perspective,'' in \emph{Proc. IEEE
  Int. Symp. High Performance Computer Architecture ({HPCA})}, 2018, pp.
  620--629.

\bibitem{jia2014caffe}
Y.~Jia, E.~Shelhamer, J.~Donahue, S.~Karayev, J.~Long, R.~Girshick,
  S.~Guadarrama, and T.~Darrell, ``Caffe: Convolutional architecture for fast
  feature embedding,'' \emph{CoRR}, vol. 1408.5093, 2014.

\bibitem{Jiang2019}
B.~Jiang, C.~Deng, H.~Yi, Z.~Hu, G.~Zhou, Y.~Zheng, S.~Huang, X.~Guo, D.~Wang,
  Y.~Song, and et~al., ``Xdl: An industrial deep learning framework for
  high-dimensional sparse data,'' in \emph{Proc. 1st Int. Workshop on Deep
  Learning Practice for High-Dimensional Sparse Data}, 2019.

\bibitem{johnson2016google}
M.~Johnson, M.~Schuster, Q.~V. Le, M.~Krikun, Y.~Wu, Z.~Chen, N.~Thorat,
  F.~Vi{\'e}gas, M.~Wattenberg, G.~Corrado, M.~Hughes, and J.~Dean, ``Google's
  multilingual neural machine translation system: enabling zero-shot
  translation,'' \emph{CoRR}, vol. 1611.04558, 2016.

\bibitem{TPU_ISCA}
N.~P. Jouppi, C.~Young, N.~Patil, D.~Patterson, G.~Agrawal, R.~Bajwa, S.~Bates,
  S.~Bhatia, N.~Boden, A.~Borchers, R.~Boyle, P.-l. Cantin, C.~Chao, C.~Clark,
  J.~Coriell, M.~Daley, M.~Dau, J.~Dean, B.~Gelb, T.~V. Ghaemmaghami,
  R.~Gottipati, W.~Gulland, R.~Hagmann, C.~R. Ho, D.~Hogberg, J.~Hu, R.~Hundt,
  D.~Hurt, J.~Ibarz, A.~Jaffey, A.~Jaworski, A.~Kaplan, H.~Khaitan,
  D.~Killebrew, A.~Koch, N.~Kumar, S.~Lacy, J.~Laudon, J.~Law, D.~Le, C.~Leary,
  Z.~Liu, K.~Lucke, A.~Lundin, G.~MacKean, A.~Maggiore, M.~Mahony, K.~Miller,
  R.~Nagarajan, R.~Narayanaswami, R.~Ni, K.~Nix, T.~Norrie, M.~Omernick,
  N.~Penukonda, A.~Phelps, J.~Ross, M.~Ross, A.~Salek, E.~Samadiani, C.~Severn,
  G.~Sizikov, M.~Snelham, J.~Souter, D.~Steinberg, A.~Swing, M.~Tan,
  G.~Thorson, B.~Tian, H.~Toma, E.~Tuttle, V.~Vasudevan, R.~Walter, W.~Wang,
  E.~Wilcox, and D.~H. Yoon, ``In-datacenter performance analysis of a {Tensor
  Processing Unit},'' in \emph{Proc. 44th Annual Int. Symp. Computer
  Architecture ({ISCA})}, 2017, pp. 1--12.

\bibitem{dragonfly}
J.~Kim, W.~J. Dally, S.~Scott, and D.~Abts, ``Technology-driven,
  highly-scalable dragonfly topology,'' in \emph{Proc. 35th Annual Int. Symp.
  Computer Architecture ({ISCA})}, 2008, pp. 77--88.

\bibitem{JKim05}
J.~Kim, W.~J. Dally, B.~Towles, and A.~K. Gupta, ``Microarchitecture of a high
  radix router,'' in \emph{Proc. 32nd Int. Symp. Computer Architecture
  ({ISCA})}, 2005.

\bibitem{BigBasin2019}
\BIBentryALTinterwordspacing
K.~Lee, ``Introducing {Big Basin}: Our next-generation {AI} hardware,'' 2019.
  [Online]. Available:
  \url{https://engineering.fb.com/data-center-engineering/introducing-big-basin-our-next-generation-ai-hardware}
\BIBentrySTDinterwordspacing

\bibitem{AngLi2019}
A.~Li, S.~L. Song, J.~Chen, J.~Li, X.~Liu, N.~Tallent, and K.~Barker,
  ``Evaluating modern {GPU} interconnect: {PCIe, NVLink, NV-SLI, NVSwitch and
  GPUDirect},'' \emph{CoRR}, vol. 1903.04611, 2019.

\bibitem{HuaweiTraining}
H.~Liao, J.~Tu, J.~Xia, and X.~Zhou, ``{DaVinci}: A scalable architecture for
  neural network computing,'' in \emph{Hot Chips 31 Symposium, Palo Alto, CA,
  USA}, 2019.

\bibitem{WaferScale}
S.~Lie, ``Wafer scale deep learning,'' in \emph{Hot Chips 31 Symposium, Palo
  Alto, CA, USA}, 2019.

\bibitem{Mayer2019}
R.~Mayer and H.-A. Jacobsen, ``Scalable deep learning on distributed
  infrastructures: Challenges, techniques and tools,'' \emph{ACM Computing
  Surveys}, vol.~53, 2019.

\bibitem{GaudiHotChips}
E.~Medina, ``Habana labs approach to scaling {AI} training,'' in \emph{Hot
  Chips 31 Symposium, Palo Alto, CA, USA}, 2019.

\bibitem{mittal2018revisiting}
R.~Mittal, A.~Shpiner, A.~Panda, E.~Zahavi, A.~Krishnamurthy, S.~Ratnasamy, and
  S.~Shenker, ``Revisiting network support for {RDMA},'' \emph{CoRR}, vol.
  1806.08159, 2018.

\bibitem{DLRM}
\BIBentryALTinterwordspacing
M.~Naumov, D.~Mudigere, H.~M. Shi, J.~Huang, N.~Sundaraman, J.~Park, X.~Wang,
  U.~Gupta, C.~Wu, A.~G. Azzolini, D.~Dzhulgakov, A.~Mallevich,
  I.~Cherniavskii, Y.~Lu, R.~Krishnamoorthi, A.~Yu, V.~Kondratenko, S.~Pereira,
  X.~Chen, W.~Chen, V.~Rao, B.~Jia, L.~Xiong, and M.~Smelyanskiy, ``Deep
  learning recommendation model for personalization and recommendation
  systems,'' \emph{CoRR}, vol. 1906.00091, 2019. [Online]. Available:
  \url{https://github.com/facebookresearch/dlrm}
\BIBentrySTDinterwordspacing

\bibitem{googRevisitingSync}
X.~Pan, J.~Chen, R.~Monga, S.~Bengio, and R.~Jozefowicz, ``Revisiting
  distributed synchronous {SGD},'' 2017.

\bibitem{FBinference}
J.~Park, M.~Naumov, P.~Basu, S.~Deng, A.~Kalaiah, D.~S. Khudia, J.~Law,
  P.~Malani, A.~Malevich, N.~Satish, J.~Pino, M.~Schatz, A.~Sidorov,
  V.~Sivakumar, A.~Tulloch, X.~Wang, Y.~Wu, H.~Yuen, U.~Diril, D.~Dzhulgakov,
  K.~M. Hazelwood, B.~Jia, Y.~Jia, L.~Qiao, V.~Rao, N.~Rotem, S.~Yoo, and
  M.~Smelyanskiy, ``Deep learning inference in facebook data centers:
  Characterization, performance optimizations and hardware implications,''
  \emph{CoRR}, vol. 1811.09886, 2018.

\bibitem{paszke2017automatic}
\BIBentryALTinterwordspacing
A.~Paszke, S.~Gross, S.~Chintala, G.~Chanan, E.~Yang, Z.~DeVito, Z.~Lin,
  A.~Desmaison, L.~Antiga, and A.~Lerer, ``Automatic differentiation in
  {PyTorch},'' 2017. [Online]. Available: \url{https://pytorch.org/}
\BIBentrySTDinterwordspacing

\bibitem{Sergeev2018}
A.~Sergeev and M.~D. Balso, ``Horovod: fast and easy distributed deep learning
  in {TensorFlow},'' \emph{CoRR}, vol. 1802.05799, 2018.

\bibitem{Shanley2002}
T.~Shanley, \emph{Infiniband Network Architecture}.\hskip 1em plus 0.5em minus
  0.4em\relax Addison-Wesley, 2002.

\bibitem{ZionHotChips}
M.~Smelyanskiy, ``Zion: Facebook next-generation large-memory unified training
  platform,'' in \emph{Hot Chips 31 Symposium, Palo Alto, CA, USA}, 2019.

\bibitem{ScaleDeep}
S.~Venkataramani, A.~Ranjan, S.~Banerjee, D.~Das, S.~Avancha, A.~Jagannathan,
  A.~Durg, D.~Nagaraj, B.~Kaul, P.~Dubey, and A.~Raghunathan, ``{ScaleDeep}: A
  scalable compute architecture for learning and evaluating deep networks,'' in
  \emph{Proc. 44th Annual Int. Symp. Computer Architecture ({ISCA})}, 2017, pp.
  13--26.

\bibitem{Alibaba-PAI}
M.~Wang, C.~Meng, G.~Long, C.~Wu, J.~Yang, W.~Lin, and Y.~Jia, ``Characterizing
  deep learning training workloads on {Alibaba-PAI},'' \emph{CoRR}, vol.
  1910.05930, 2019.

\bibitem{IntelSpringCrest}
A.~Yang, N.~Garegrat, C.~Miao, and K.~Vaidyanathan, ``Deep learning training at
  scale: {Spring Crest} deep learning accelerator,'' in \emph{Hot Chips 31
  Symposium, Palo Alto, CA, USA}, 2019.

\bibitem{TPUhotchips}
C.~Young, ``Evaluation of the {Tensor Processing Unit}: A deep neural network
  accelerator for the datacenter,'' in \emph{Hot Chips 29 Symposium, Palo Alto,
  CA, USA}, 2017.

\bibitem{oam_initiative}
\BIBentryALTinterwordspacing
W.~Zhao, ``{OCP Accelerator Module (OAM)},'' 2019. [Online]. Available:
  \url{https://engineering.fb.com/data-center-engineering/accelerator-modules/}
\BIBentrySTDinterwordspacing

\bibitem{Zheng2020}
Q.~Zheng, B.-Y. Su, J.~Yang, A.~Azzolini, Q.~Wu, O.~Jin, S.~Karandikar,
  H.~Lupesko, L.~Xiong, and E.~Zhou, ``Shadowsync: Performing synchronization
  in the background for highly scalable distributed training,'' \emph{CoRR},
  vol. 2003.03477, 2020.

\bibitem{Zinkevich2010}
M.~Zinkevich, M.~Weimer, L.~Li, and A.~J. Smola, ``Parallelized stochastic
  gradient descent,'' in \emph{Advances in Neural Information Processing
  Systems ({NIPS}) 23}.\hskip 1em plus 0.5em minus 0.4em\relax Curran
  Associates, Inc., 2010, pp. 2595--2603.

\end{thebibliography}

\end{document}